\documentclass{aa}

\usepackage{txfonts}

\usepackage[english]{babel}

\usepackage{graphicx}

\usepackage{euscript}
\newcommand{\Fcurve}{\mathfrak{F}}

\begin{document}

\title{
Low-rate accretion onto isolated stellar-mass black holes
}

\author{G.M.~Beskin\inst{1,2} \and S.V.~Karpov\inst{1,2}}
\authorrunning{G.M.~Beskin  \and S.V.~Karpov}
\titlerunning{Low-rate accretion}
\offprints{G. Beskin,
\email{beskin@sao.ru}}

\institute{
Special Astrophysical Observatory, Nizhnij Arkhyz,
              Karachaevo-Cherkesia, 369167, Russia \\
\and Isaac Newton Institute of Chile, SAO Branch \\
}

\date{Received / Accepted  }

\abstract{
Magnetic field behaviour in a spherically-symmetric accretion flow for
parameters typical of single black holes in the Galaxy is discussed.
It is shown that in the majority of Galaxy volume, accretion onto single
stellar-mass black holes will be spherical and have a low accretion rate
($10^{-6} - 10^{-9}$ of the Eddington rate).
An analysis of plasma internal energy growth during the infall
is performed. Adiabatic heating of collisionless accretion flow due to
magnetic adiabatic invariant conservation is $25\%$ more efficient than in
the standard non-magnetized gas case.
It is shown that magnetic field line reconnections in discrete current sheets
lead
to significant nonthermal electron component formation. In a framework of
quasi-diffusion acceleration, the "energy-radius" electron distribution is
computed and the function describing the shape of synchrotron radiation
spectrum is constructed. It is shown that nonthermal electron emission leads
to formation of a hard (UV, X-ray, up to gamma), highly variable spectral
component in addition to the standard
synchrotron optical component first derived by Shvartsman
generated by thermal electrons in the magnetic field of accretion flow. 
For typical interstellar medium parameters, a black hole at 100 pc distance
will be a 16-25$^{\rm m}$ optical source coinciding with the highly variable 
bright X-ray counterpart, while the variable component of optical emission will
be about 18-27$^{\rm m}$. The typical time scale of the variability is
$10^{-4}$ sec, with relative flare amplitudes of 0.2-6\% in various spectral bands.
Possible
applications of these results to the problem of search for single black holes
are discussed.
\keywords{accretion -- black hole physics -- Galaxy: stellar content --
ISM: general -- magnetic fields -- plasmas -- X-rays: general }
}

\maketitle

\section{Introduction}
\label{sec_introduction}

Even though more than 60 years have passed since the theoretical
prediction of black holes as an astrophysical objects
(Oppenheimer \& Snyder \cite{oppenheimer}) in some sense they have not been discovered yet.
To identify an object as a black hole, one needs to show
that its mass exceeds $3M_{\odot}$, its size is close to
$r_g = 2G M/c^2$ and it has an event horizon instead of a normal surface --
the distinguishing property of black holes which
separates them from massive compact objects of finite size in some theories
of gravity (Will \cite{will}). However, only the two former criteria are used now
for selection of black hole candidates of two types:
a) with masses of 5-18 $M_{\odot}$, in X-ray binaries (see, for example,
Greiner et al. \cite{greiner});
and b) supermassive black holes in galaxy nuclei with masses of
$10^6 - 10^{10} M_{\odot}$ (Shields \cite{shields_1999}).
Existence of the event horizon in such objects is usually implied by the
absence of periodic pulsations of the X-ray emission from strong
regular magnetic fields (the black hole "no-hair" theorem) and I type X-ray
flares due to thermonuclear bursts of the accreted matter on the surface of
the neutron star. At the same time, typical masses of X-ray pulsars and
bursters are close to the typical neutron star value of 1.4 $M_{\odot}$ while
black hole candidates, missing pulsations and X-ray flares, have masses of
5-18 $M_{\odot}$ (Miller et al. \cite{miller_1998}). The absence of an event horizon in
low-mass objects is not a proof of its existence in higher-mass ones.

High accretion rates in X-ray binaries and active galactic nuclei
result in the screening of regions close to the event
horizon, and the most luminous parts of accretion flow are situated at distances of
$10 - 100 r_g$ (Chakrabarti \cite{chakrabarti}; Cherepashchuk \cite{cherepashchuk}) 
where general relativity effects are negligible.

There is a very effective way to get information about the innermost parts of accretion disks 
in X-ray binaries as well as AGNs -- the investigation of the broad (and sharp) iron $K_{\alpha}$
fluorescent emission line (see review by Reynolds \& Nowak \cite{reynolds_2003}).
Its intensity and shape depend on the accreted plasma distribution and behaviour until the last 
stable orbit (0.62 $r_g$ for a extremely spinning Kerr black hole and 3 $r_g$ for a Schwarzschild one)
(Miller et al. \cite{miller_2004}; Miniutti et al. \cite{miniutti}). However since the photons generated
at the different distances from the horizon are mixed in the line profile it is not
possible to extract the manifestations of gravitational fields close to the horizon only.
This may be possible by study of variability of the iron line
(Reynolds et al. \cite{reynolds_2004}).

At the same time, single stellar-mass black holes, which accrete interstellar
medium of  low density ($10^{-2} - 1 $cm$^{-3}$), are the ideal case for
detection and study of the event horizon. Shvartsman (\cite{shvartsman_1971}) first demonstrated
that an emitting halo of accreted matter forms around such objects and
generates optical featureless emission. The majority of such emission comes
from the regions near the horizon at $(3 -- 5) r_g$. Spherical accretion onto
the single stellar-mass black holes has been studied in detail in the
works of several authors (Bisnovatyi-Kogan \& Ruzmaikin \cite{bisnovatyi_1974}; Meszaros \cite{meszaros};
Ipser \& Price \cite{ipser_1977, ipser_1982}) and the main conclusions of Shvartsman have been
confirmed.

The most striking property of the accretion flow onto the single black hole is
its inhomogeneity -- the clots of plasma act as a probe testing
the space-time properties near the horizon. The characteristic timescale of
emission variability is $\tau_v\sim r_g/c\sim 10^{-4} - 10^{-5}$ sec
and such short stochastic variability may be considered as a distinctive
property of black hole as the smallest possible physical object with a given mass.
Its parameters -- spectra, energy distribution and light curves --
carry important information on space-time properties of the horizon
(Beskin \& Shvartsman \cite{beskin_1976}). 

The general observational appearance of a single
stellar-mass black hole at typical interstellar medium densities is the same
as other optical objects without spectral lines -- DC-dwarfs and ROCOSes (Radio
Objects with Continuous Optical Spectra, a subclass of blazars)
(Beskin \& Mitronova \cite{beskin_1991}, Pustilnik \cite{pustilnik_1977}, Shvartsman \cite{shvartsman_1977}; Beskin et al. \cite{beskin_2000}). 
The suggestion that isolated BHs can be among them is the basis of the observational programme of search
for isolated stellar-mass black holes -- MANIA (Multichannel Analysis of
Nanosecond Intensity Alterations).
It uses photometric observations of candidate
objects with high time resolution, special hardware and data analysis
methods (Shvartsman \cite{shvartsman_1977}; Beskin et al. \cite{beskin_1997}).

In observations using the 6-meter telescope of the Special Astrophysical Observatory of 40 DC-dwarfs
and ROCOSes, only upper limits for variability levels of 20\% -- 5\% on the timescales of $10^{-6}$ -- 10 sec,
respectively, were obtained, i.e. BHs were not detected (Shvartsman et al. \cite{shvartsman_1989a}, 
\cite{shvartsman_1989b}, Beskin et al. \cite{beskin_2000}).

Recently, some evidences appeared that single stellar-mass black holes may
be found among the stationary unidentified gamma-ray sources (Gehrels et al.
2002), gravitational lenses causing long-lasting MACHO
events (Bennett et al. 2001) and white dwarf -- black hole binaries
detected by means of self-microlensing  flashes (Beskin \& Tuntsov \cite{beskin_2002b}).
In the last case, the mass transfer from the white dwarf is absent and a
black hole behaves as a single one.

In the present work we study spherical accretion onto a single stellar-mass
black hole at low accretion rates of $10^{8} - 10^{13}$ g/s.
This corresponds to the range of interstellar medium densities of
$0.002 - 0.1$cm$^{-3}$ and a 10 $M_{\odot}$ object moving with a
velocity of 20--40 km/s (Bondi \& Hoyle \cite{bondi_1944}). This is true for about 90\%
of the Galaxy volume (McKee \& Ostriker \cite{mckee}).

Spherical accretion with equipartition of energies, i.e. with roughly equal
densities of the magnetic and kinetic energies of plasma has been considered
in many papers (Shvartsman \cite{shvartsman_1971}; Bisnovatyi-Kogan \& Ruzmaikin \cite{bisnovatyi_1974};
Kowalenko \& Melia \cite{kowalenko}; Ipser \& Price \cite{ipser_1977,ipser_1982}).
The uniqueness of our approach is in taking into account the significantly
non-thermal nature of electron energy distribution function (its synchrotron emission determines
the appearance of a black hole).
It may be roughly considered as a superposition of two components (this
approach is known as "hybrid plasma", see Coppi (\cite{coppi}) and references
therein) -- thermal electrons and accelerated electron beams, formed in
current sheets where magnetic energy is dissipated in a way similar to solar
flares (Pustilnik \cite{pustilnik_1978,pustilnik_1997}). The latter process supports the
equipartition of energies.
As a result, the emission of the accretion flow consists
of a quasi-stationary "thermal" part with a wide-band spectrum from infrared
to ultraviolet, and a highly variable flaring nonthermal component. Each
such flare is generated due to the motion of the accelerated electron beam
in the magnetic field. Its light curve carries information on
the magnetic and gravitational field structure near the black hole horizon.
Nonthermal luminosity reaches several percents of the total luminosity and
may even exceeds it at low rates, while its spectrum covers
spectral bands from optical to hard X-ray. This result leads to possible
modifications of the search strategy.

In \S 2, the main characteristics of accretion onto single black holes in the
Galaxy are discussed.

In \S 3, the electron distribution function in phase space is built, in
\S 4, the thermal and nonthermal component luminosities are determined, and
in  \S 5 the shape of its spectra is studied.

In \S 6, the temporal behaviour of single electron beam emission is studied
and some conclusions on the variability of accretion flow are made.

In \S 7, the main results of this work are summarized, and in \S 9, possible
directions of future work are discussed.

\section{Nature of accretion and basic parameters of the model}
\label{sec_model}

\subsection{Accretion rate for different parameters of black holes and interstellar medium}

Contemporary models of massive star evolution, its dynamics and data on
black hole candidates in X-ray binaries and microlensing events suggest
that the most probable mass $M$ of a single black hole is
$10 M_{\odot}$(Greiner et al. \cite{greiner}, Fryer \& Kalogera \cite{fryer},
Agol \& Kamionkowski \cite{agol_2002b}), while
its velocity is in the $10-50$ km/s range and the gas capture cross-section is
defined by the Bondi radius ${r_c}$
(Bondi \& Hoyle \cite{bondi_1944}):
\begin{equation}
r_c = \frac{2GM}{V^2 + c_{s}^2}\; ,
\end{equation}
where $c_s$ is the  sound speed in the interstellar medium.

The interstellar medium consists of at least three components
(McKee \& Ostriker \cite{mckee}) -- cold ($T\sim 10^{2} K$) and dense
($n\sim 10^2 $cm$^{-3}$) neutral hydrogen clouds, warm partly ionized hydrogen
($n\sim 0.1-0.2$ cm$^{-3}, T\sim 10^4 $K) ones and fully ionized coronal hydrogen
($n\sim 0.002 $cm$^{-3}, T\sim 10^6 $K) clouds. Motion may be
subsonic as well as supersonic -- the sound speed changes in the
$1.5 -- 150$ km/s range.

For the black hole moving supersonically in a uniform medium, an accretion
rate is determined by the behaviour of matter behind the gravitation centre,
where the tangential gas velocity component vanishes in the shock wave and the gas
falls towards the black hole in a wide cone(Bondi \& Hoyle \cite{bondi_1944}, Shvartsman
\cite{shvartsman_1971}, Illarionov \& Sunyaev \cite{illarionov}, Font \& Ibanez \cite{font}).
For the typical case of interstellar medium that is collisionless
on scales of the capture radius, a mode of accretion from a very thin "tail",
where the tangential velocity vanishes due to the magnetic field, is possible.
The thickness of such a tail may be estimated as

\begin{equation}
\frac{r_{tail}}{r_c}\approx\left(\frac{B_{\infty}^2/8\pi}{\rho V^2/2}\right)^
{1/2}\approx\frac{c_s}{V},
\end{equation}

where $B_{\infty}$ is the magnetic field strength at infinity, 
$V$ is the black hole velocity and rough equipartition of
thermal and magnetic energies in the interstellar medium is taken into
account. Accretion flow is continuous due to
interstellar magnetic fields -- the proton Larmor radius is $10^7 -- 10^8$ cm which
is at least $10^6$ times smaller than $r_c$. The interstellar magnetic field in the Galaxy
($10^{-4} -- 10^{-5}$ G) is frozen-in.
The accretion rate for such a regime is determined by the expression
(Bondi \& Hoyle \cite{bondi_1944})

\begin{equation}
\dot M = \frac{4\pi G^2 M^2\rho}{(V^2 + c_s^2)^{3/2}}\; ,
\label{eqn_bondi_rate}
\end{equation}
where $\rho$ in the interstellar medium density.

Using the expression for the Eddington accretion rate
$\dot M_{edd} = L_{edd}/c^2 =\frac{2\pi m_p cr_g}{\sigma_T}=1.4\cdot 10^{18}
(M/10M_{\odot})$ g/s, where $m_{p}$ is
the proton mass, $\sigma_T$ is the Tompson cross-section,
we have a useful normalization, measuring
the mass in units of $10M_{\odot}$, density in $1$ cm$^{-3}$, and velocity
in $16$ km/s

\begin{equation}
\dot m = \dot M/\dot M_{edd} = 1.3\cdot 10^{-5}M_{10}n_{1}(V^2 + c_{s}^
{2})_{16}^{-3/2}\; .
\label{eqn_dotm}
\end{equation}

\begin{figure}
\center
{\centering \resizebox*{1\columnwidth}{!}{\includegraphics{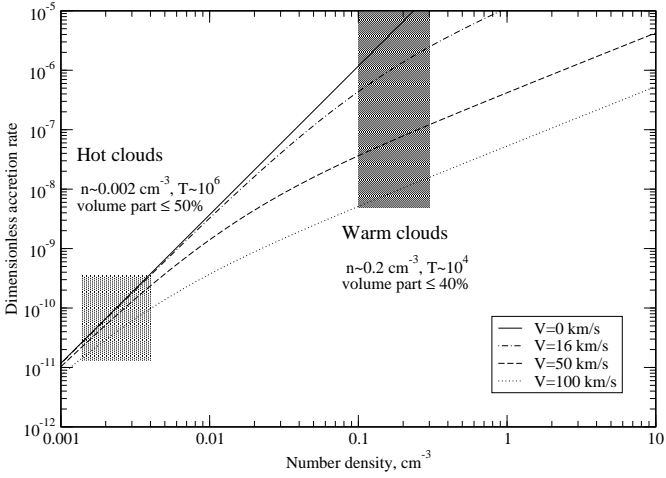}} \par}
\caption{Dimensionless accretion rates $\dot m =\dot M c^2/L_{edd}$
as a function of interstellar medium density for various black hole
velocities.}
\label{fig_accretion_rate}
\end{figure}

In Fig.\ref{fig_accretion_rate}, the dependence of $\dot m$ on $n$
is shown for various velocities of the  black hole motion (the dependence of
the sound speed $c_s$ on the temperature $T$ due to hydrostatic equilibrium
of the ISM is also taken into account). Only in the cold clouds the accretion
rate may reach the Eddington level, but due to the relative rareness of such
clouds (5\% of the Galactic volume) this case is improbable. Note however that
a black hole may initially be born in such a cloud, and so become a very
bright source with a luminosity up to $10^{38} -- 10^{40}$ erg/s (and so,
be a so-called "ultraluminous" source (Roberts et al. \cite{roberts}), like those
observed in other galaxies). For the warm hydrogen, whose volume fraction
$\approx 40\%$, the  accretion rate will be $10^{-6} -
10^{-8}$, while for hot hydrogen ($\approx 50 \%$ volume) it will be significantly lower. 
Later, we will assume such cases as the most typical for
single black holes in the Galaxy.

\subsection{The role of interstellar medium inhomogeneities. Spherical or disc-like accretion?}

The observational appearance of the accreting black holes depends crucially
on the regime of matter flow near the relativistic object. So, in binaries
the captured angular momentum of matter is high enough, and a disc-like
accretion regime takes place (Shakura \& Sunyaev \cite{shakura}). For the case of
accretion from interstellar medium, however, this problem is more
complicated.

In the recent series of papers on the analysis of accretion onto single
black holes, it was reasoned that a disc-like regime is realized for such a
case also (Fujita et al. \cite{fujita}; Agol \& Kamionowski \cite{agol_2002b}; Chisholm et al \cite{chisholm}).
However, it seems they used overestimated values for the captured angular
momentum.

Davies \& Pringle (\cite{davies}) consider analytically the problem of accretion onto the moving
gravitational centre of non-uniform interstellar medium
(inhomogeneities of density as well as of velocity are considered) and show
that, in the first order of inhomogeneities, the magnitude accretion rate is
still described by the Bondi-Hoyle formula (\ref{eqn_bondi_rate}), and the captured
angular momentum is zero.

The latter result may easily be understood by considering a simple model
of accretion from the tail in which the traversal velocity component vanishes.
This tail, on which the angular momentum is zero by definition, will not, in
general, be a straight line, and the capture cross-section will not be
a circle, and so, the total captured angular momentum will become zero too.
In a more realistic picture, however, the role of gas pressure is important,
the matter accretes in a wide cone, and the total captured angular momentum
may be significant, although, as it was shown by numerical simulation
results of Sawada et al. (\cite{sawada}), Ruffert (\cite{ruffert_1997,ruffert_1999}),
it is much smaller than the
usually-used estimation of Illarionov \& Sunyaev (\cite{illarionov}):
\begin{equation}
l_m = \frac{1}{4}\beta V_K r_c\; ,
\end{equation}
(where $\beta = \frac{\Delta\rho}
{\rho},\; \frac{\Delta V}{V}$ 
-- relative density and velocity variations on ${2r_c}$ scale, 
$V_K =\sqrt{\frac{V^2 +c_s^2}{2}}$ -- Keplerian speed at the capture radius).
So, this estimation may be considered as an upper limit in a
very narrow class of objects.

Detailed analysis and numerical simulations leads to the more realistic
expression for the captured angular momentum (Ruffert \cite{ruffert_1997,ruffert_1999}):
\begin{equation}
l\sim 0.1l_m\; .
\end{equation}
For the spherical accretion regime the specific angular momentum of the
captured matter must be smaller than that of the black hole last stable
orbit, i.e. $l < \sqrt{3}cr_g$ for Schwarzschild metric.
Then
\begin{equation}
\beta < 40\sqrt{6}\frac{\sqrt{V^2 + c_s^2}}{s}.
\end{equation}

Due to the characteristic dispersion scale in the interstellar medium
being of the order of $(\Delta V)^2 = 1.1(r/1\mbox{pc})^{0.76}(\frac{\mbox{km}}{\mbox{s}})^2$
(Larson \cite{larson}; Falgarone \& Phillips \cite{falgarone}), then
\begin{equation}
1.1(\frac{2r_c}{1pc})^{0.38} < 40\sqrt{3}\frac{V_0^2}{s},
\end{equation}
where $V_0 = \sqrt{V^2 + c_s^2}$. Finally
\begin{equation}
V_0 > 17 M_{10}^{0.138} \mbox{ km/s }\; .
\end{equation}

Therefore for nearly any black hole velocity, interstellar medium turbulent
motion cannot prevent realization of the spherical accretion regime.
\footnote{As already noticed, an obvious exception is the case of cold
clouds where sonic speed is low and at $v < 16-17$ km/s a disk may form. We
are planning to discuss this case in a separate paper.}
Furthermore, density fluctuations cannot prevent it either.
For $\frac{\Delta\rho}{\rho}\sim (\frac{r}{1pc})^{\frac{11}{6}}$ (Armstrong
et al. \cite{armstrong}), it may be easily shown that
\begin{equation}
V_0 > 3.7 M_{10}^{0.39} \mbox{ km/s }\; .
\end{equation}

It is clear that even in cold clouds, density fluctuations cannot lead to the
disc accretion regime.

\subsection{Radial structure of the flow}
\label{sec_model_structure}

The general solution of hydrodynamical problem of accretion onto a non-moving gravitating
center derived by Bondi (\cite{bondi_1952}) determines radial profiles of various
accretion flow parameters as a function of the distance to the black hole,
accretion rate and gas adiabatic index. It is important for us that
for $\gamma<5/3$ (that is always true for interstellar gas) there is a 
 "sonic point" in the flow, passing which the gas motion becomes
supersonic, and the gas velocity near the gravitating center has an
asymptotic behaviour $v\propto r^{-1/2}$ (and so $\rho\propto r^{-3/2}$
for density). In an approximate description, we may extrapolate
such a behaviour to the capture radius scale and use it for the whole flow.

In the black hole tail, where the matter flow stops, the thermal,
gravitational and magnetic energy densities become nearly equal (since in the
collisionless case, the matter is stopped due to the magnetic pressure and
plasma oscillations (Illarionov \& Sunyaev \cite{illarionov})). This equipartition of
energies (at least magnetic and gravitational ones) is preserved in the following
infall (so-called Shvartsman equipartition theorem (Shvartsman \cite{shvartsman_1971})),
so we may assume that (Bisnovatyi-Kogan \& Ruzmaikin \cite{bisnovatyi_1974})
\begin{equation}
\frac{B^2}{8\pi}=\frac12\rho v^2=\alpha^2 \frac{GM\rho}{r} \mbox{  ,}
\end{equation}
where $\alpha^2\approx1/3$ (which corresponds to equal amounts of
gravitational energy transition into kinetic, magnetic and gravitational ones).
Therefore, for the parameters of the accretion flow we have

\begin{equation}
v = \alpha c\sqrt{\frac{r_g}{r}} = \alpha c R^{-1/2},
\label{eqn_radial_v}
\end{equation}
\begin{equation}
\rho = \frac{\dot M}{4\pi r^2 v} = \frac{m_p \dot m}{2\sigma_T \alpha r_g}
R^{-3/2},
\label{eqn_radial_rho}
\end{equation}
\begin{equation}
\frac{B^2}{8\pi} = \frac12 \rho v^2 = \frac{\alpha m_p c^2 \dot m}{4\sigma_T
r_g}R^{-5/2},
\label{eqn_radial_B}
\end{equation}
where the dimensionless values for radius $R=r/r_g$ and the accretion rate
from equation (\ref{eqn_dotm}) are used. Numerical values of these parameters
are
\begin{equation}
n = \frac{\rho}{m_p} = 4.33 \cdot 10^{12} \ \dot m_{-5} M_{10}^{-1} R^{-3/2}
\mbox{ cm$^{-3}$},
\end{equation}
\begin{equation}
B = 8 \cdot 10^{4} \ \dot m^{1/2}_{-5} M_{10}^{-1/2} R^{-5/4} \mbox{ Gauss}.
\end{equation}

The magnetic field has a quasi-radial sectorial structure (the radial component grows much
faster than the tangential one, which is proportional to the square root of
distance; magnetic field lines are stretched).

The assumption introduced earlier on the equipartition of magnetic, kinetic
and gravitational energy requires the existence of 
magnetic flux (and, therefore magnetic energy) dissipation mechanisms.
It was first noted by Shvartsman (\cite{shvartsman_1971}); their possible observational
appearances were discussed by Illarionov and Sunyaev (\cite{illarionov}); a
model-independent estimation of the energy dissipation rate was
proposed by Bisnovatyi-Kogan \& Ruzmaikin (\cite{bisnovatyi_1974}) and from a different point of view
by Meszaros (\cite{meszaros}); alternative approaches were discussed by Scharlemann (\cite{scharlemann})
and Kowalenko \& Melia (\cite{kowalenko}).

The dissipation rate of such a mechanism may be estimated as follows
(Bisnovatyi-Kogan \& Ruzmaikin \cite{bisnovatyi_1974}). Scaling laws for the magnetic field in a given volume
element for the frozen-in field and equipartition are correspondingly
\begin{equation}
\left(\frac{d}{dt}\frac{B^2}{8\pi}\right)_{frozen-in}=-4\frac vr
\frac{B^2}{8\pi} \mbox{  ,}
\end{equation}
\begin{equation}
\left(\frac{d}{dt}\frac{B^2}{8\pi}\right)_{equipartition}=
-\frac52\frac vr\frac{B^2}{8\pi} \mbox{  ,}
\end{equation}
\begin{equation}
\frac{B^2}{8\pi}=\frac12\rho v^2=\alpha^2 \frac{GM\rho}{r} \mbox{  .}
\end{equation}
For the preservation of equipartition state, the power equal to the
difference of these expressions must be dissipated in a volume element:
\begin{equation}
\frac{d E}{dV dt}=\frac32\frac{v}{r}\frac{B^2}{8\pi},
\label{eqn_dissipation_rate}
\end{equation}
or the same for a spherical shell
\begin{equation}
\frac{d E}{dR dt}=\frac{3\alpha^2}{4}\frac{\dot M c^2}{R^2},
\label{eqn_dissipation_radial}
\end{equation}
which, by integrating over the whole volume, gives
\begin{equation}
\frac{dE}{dt}=\frac34\alpha^2\dot Mc^2=\frac14\dot Mc^{2},
\label{eqn_dissipation_total}
\end{equation}
which means that in the case of equipartition for the spherical flow
described earlier, as much as 25\% of infalling matter rest energy is released
through this dissipation mechanism only. This leads to an additional
super-adiabatic heating of the gas (Bisnovatyi-Kogan \& Ruzmaikin \cite{bisnovatyi_1974};
Meszaros \cite{meszaros}; Ipser \& Price \cite{ipser_1977}), and unavoidably changes the temperature
radial profile and the luminosity of the accretion flow.

\subsection{Flaring dissipation of the magnetic field and electron
acceleration in the current sheets}
\label{sec_acceleration}

It is clear from simple physical reasons that dissipation of the magnetic
field means that it is no longer frozen-in, i.e. the mean conductivity becomes
much lower and relative motion of the magnetic field and plasma
appears (the mean diffusion time becomes comparable to the free-fall one),
currents begin to flow and heat the gas. These processes may take place
either continuously in turbulent accretion flow as an Ohmic dissipation of the magnetic field (Bisnovatyi-Kogan \&
Ruzmaikin \cite{bisnovatyi_1974}; Bisnovatyi-Kogan \& Lovelace \cite{bisnovatyi_1997}) or in compact
enough separate regions. In the latter case, the dissipation process has a highly
discrete nature (this possibility was noted by Bisnovatyi-Kogan \& Lovelace
(\cite{bisnovatyi_2000})) of single events of magnetic field line reconnections.

In the present work we consider the latter case. It seems to be more
realistic due to the fact that Alfven velocity which determines the speed of
energy exchange between such a reconnection region and the surrounding plasma is
nearly equal to the free-fall velocity at which a high magnetic field
gradient forms. This is analogous to the case of continuous energy
supply to the magnetic field inhomogeneities in the solar corona and its
flaring (discrete) dissipation and conversion to acceleration of particles
and anomalous turbulent heating. This is the case of formation of turbulent
current sheets, which leads to fast magnetic field line reconnection,
avalanche-like growth of energy release (the flare itself) and threshold
switch-off of energy dissipation processes (Sweet \cite{sweet}; Petchek \cite{petchek};
Parker \cite{parker}; Spitzer \cite{spitzer}; Syrovatskii \cite{syrovatsky}). Multi-frequency observations of
solar flares support this mechanism. On the other hand, the similarity
of statistical properties of flaring activity of the Sun (Lu \& Hamilton \cite{lu_1991};
Lu et al. \cite{lu_1993}) and UV Cet stars (Gershberg \cite{gershberg}), X-ray binaries (Kawaguchi \&
Mineshige \cite{kawaguchi}) gamma-ray  bursts and active galactic nuclei argues in favor
of the universality of such processes. So, we may assume that the main mechanism
providing magnetic energy dissipation in the accretion flow is a reconnection
of the magnetic field lines in the current sheets (regions of high magnetic
field gradient).

Energy dissipation in the current sheet itself may be considered as a simple
Joule heating $Q = jE = j^{2}/\sigma$, where $j=(c/4\pi)\cdot \mbox{rot}
\vec B\propto\frac{\Delta B} {a}$ -- current density, $a$ -- current sheet
thickness, $\sigma$ -- some effective conductivity. When
current reaches some threshold value, $j_c = en_c c_i$, where $n_c$ --
electron number density, $c_i$ -- ion sound speed, it generates
strong ion-acoustic turbulence that lowers the conductivity by 9-10 orders of
magnitude (electrons begin to collide with plasmons and heat them). The
analysis of particle acceleration in such a current sheet (which may be
considered as a superposition of direct electric field acceleration and
diffusion -- elastic scattering on plasmons) was performed in 
Pustilnik (\cite{pustilnik_1978, pustilnik_1997}). In the  framework of this model,
the accelerated
particle energy distribution (electrons mostly) naturally appears to be a
power-law up to energies large enough (Pustilnik \cite{pustilnik_1978}) and has a shape
\begin{equation}
f_0(\gamma)=\frac{1}{\Gamma}\left(\frac{\Gamma}{\gamma}\right)^3e^
{-\frac{\Gamma}{\gamma}},
\label{eqn_distribution_ejected}
\end{equation}
where $\Gamma$ corresponds to the mean energy of a particle in a beam.

We estimate the maximal energy a electron reaches by acceleration in such
current sheet. By using expressions for gas density and magnetic field
strength (\ref{eqn_radial_rho}),(\ref{eqn_radial_B}) for low accretion
rates ($\dot m < 10^{-5}$), we have for the event horizon
$n\le 4\cdot 10^{12}$ cm$^{-3}$ and $B_g\le 1.7\cdot10^5$ Gs. Near $r_g$
velocities of electrons and plasmons are nearly equal to the speed of light,
and so, the maximum current density is $j_{max} = enc$, and the maximum
effective field is $E_{max} =\frac{enc}{\sigma^{*}}$, where $\sigma^{*}$ is
anomalous resistivity; for our case $\sigma^{*}\sim\frac{10^2}{4\pi}
\omega_{oe}$, where $\omega_{oe}$ -- Lengmure frequency ($\omega_{oe} =
5.65\cdot 10^4\cdot n^{1/2}$). Therefore, the maximum gamma-factor is
\begin{equation}
\gamma_{max}\sim\frac{eE_{max}r_g}{m_e c^2}\sim \frac{e^2 nr_g}
{\sigma^{*} m_e c}\sim\frac{e^2 r_g n^{1/2}}{4.5\cdot 10^5 m_e c}\sim 10^{5}.
\end{equation}

It is very difficult to estimate the fraction $\xi$ of the dissipated
energy carried away by the accelerated particles -- electrons during the 
acceleration process generate ion-acoustic and Lengmure plasma oscillations.
Also, topology changes in reconnections lead to global matter motions.
A study of such processes in the solar flares shows that particles (electrons
mostly) carry away from 10\% to 50\% of energy stored in the magnetic
field inhomogeneities (Hudson \& Ryan \cite{hudson}). We assume $\xi = 0.1$ as a
reasonable lower limit for further estimations. The ($1-\xi$) fraction of the
dissipated energy goes into the surrounding plasma heating. In the
collisionless case this is due to generation of MHD turbulence in
the current sheet. Also, as the speed of plasmons is nearly equal to the Alfven
speed, which in turn is equal to the free-fall velocity for the equipartition
case, the heating is nearly uniform through the whole accretion flow.
It is also difficult to determine the motion of the accelerated particle beam.
Of course, electrons are moving along the magnetic field lines, tracing its
topology. Its motion is local due to the relative smallness of Larmor radius in
comparison to the characteristic scale of the accretion flow, Schwarzschild
radius $r_g$ 
\begin{equation}
\left(\frac{r_L}{r_g}\right)^2=\frac{\gamma^2}{\alpha\dot m}\frac{m_e}{m_p}\sqrt{\frac{2\sigma_T}{3\pi r_g^2}} \ll 1\mbox{ .}
\end{equation}

For a typical single stellar-mass black hole accretion rates,
electron-electron and electron-ion energy exchange is highly ineffective
(this question has been discussed in detail in the framework of advective
disk models, see  Mahadevan \& Quataert (\cite{mahadevan}) and Bisnovatyi-Kogan \& Lovelace (\cite{bisnovatyi_2001})),
so we may neglect collisional energy losses for the accelerated particles.

Furthermore, in the case of a Maxwellian distribution of the background plasma
electrons and moderate mean gamma-factors of the accelerated nonthermal
electrons ($10-10^2$ of the mean thermal one), the total electron
distribution $f(\gamma)$ always (for energies above thermal peak) satisfies
the inequality $\frac{df(\gamma)}{d\gamma} < 0$, and so is stable to the
generation of plasmons with wave vectors parallel to the motion direction
(Kaplan \& Tsytovich \cite{tsytovich}). For the same reason, all generated instabilities
vanish rapidly and the beam is stabilized. However, the induced generation of
plasmons with non-collinear wave vectors is possible; this leads to the
dissipation of the beam with a characteristic time of $t_{dis}\sim 10^2
\frac{n<\gamma>}{n_{*}\omega_{oe}}$, where $n_{*}$ is the electron density
of the beam (Kaplan, Tsytovich \cite{tsytovich}). It is easily seen that
$\frac{t_{dis}}{t_{ff}}\sim r^{-3/4}$, so, the beam born near $r_g$ has more
chance of surviving. Also it is seen that for $\frac{n}{n_{*}} > 10^2$ the
beam has not enough time to dissipate in the free-fall time scale, so, we
may neglect this scattering and thermalization of the beam and consider it
as a system of non-interacting electrons in the external magnetic field.

\begin{figure}
{\centering \resizebox*{1\columnwidth}{!}{\includegraphics{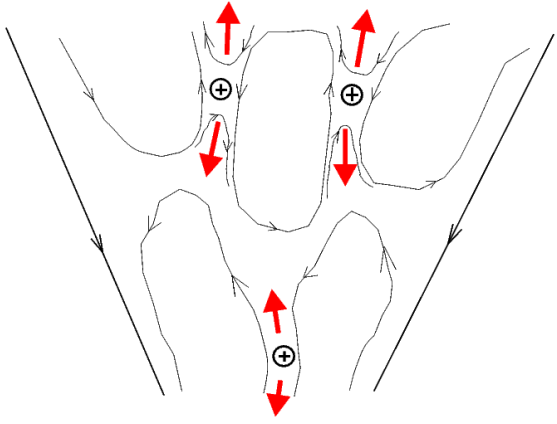}} \par}
\caption{A schematic picture of the accretion flow structure. Reconnection
regions and ejected plasma flows are shown. The direction of the particles
acceleration (current flow) is perpendicular to the viewing plane.
}
\label{fig_current_sheet}
\end{figure}

The magnetic field dissipation in the current sheet is accompanied
by significant dynamical effects; the reconnected magnetic field lines are
ejected through the current sheet ends carrying away the gas that
lowers the current sheet density (see Fig. \ref{fig_current_sheet}). 

The results of the solar corona and
chromosphere observations (Dere \cite{dere}; Innes et al. \cite{innes}) show that the
reconnection speed is often significantly less than the Alfven one
($\approx 0.1V_A$). In such a case the density and temperature fluctuations
may not have enough time to vanish, and this may lead to additional
variability of the accretion flow.

The magnetic energy dissipation mechanism proposed here is not a unique one.
However, it is in a good agreement with the observations of the Sun
(Pustilnik \cite{pustilnik_1997}). Also, it provides an opportunity for interpretation of
the universal energetic spectrum of flares on various objects -- from active
galactic nuclei to UV-Cet stars in the framework of a single mechanism.

To complete the picture, we briefly review other models.
Fast reconnections leading to particle acceleration were considered by
Lazarian \& Vishniac (\cite{lazarian_1999, lazarian_2000}), the role of various instabilities in
creation of anomalous resistivity and particles acceleration has been
considered by Birk et al. (\cite{birk}), and processes of particle acceleration
in various astrophysical objects by Bisnovatyi-Kogan \& Lovelace
(\cite{bisnovatyi_1997, bisnovatyi_2000, bisnovatyi_2001}).

The dissipated magnetic
energy is converted mostly to acceleration of electrons. In the framework of
our paper this means that for analysis of emission processes we may take into
account the electron component only.

\section{Derivation of electron distribution}
\label{sec_distribution}

\subsection{Note on adiabatic heating}
\label{sec_adiabatic}

At the low accretion rates considered, the gas is fully collisionless,
and so the particle energy distribution is determined by superposition
of three different factors -- adiabatic heating, synchrotron cooling and 
nonthermal heating due to particle acceleration in the current sheets.
The former dominates at large distances, and so its precise treatment
is very important.

At the level of single particle motion, adiabatic heating of collisionless
magnetized gas is due to the conservation of the adiabatic invariant
(Landau \& Lifshitz \cite{landau})
\begin{equation}
I=\frac{3cp_t^2}{2eB}
\end{equation}
of the charged particle in the magnetic field evolving slowly on a timescale
of a single Larmor revolution. This basically means the conservation of the
phase volume element per particle  $p_t^3\cdot V=const$. Here we neglect the
motion parallel to the magnetic field line -- only the perpendicular momentum $p_t$
grows due to compression, while the parallel one $p_{||}$ is determined mostly by
the initial conditions. 

Note that classical relation between the spatial part of the phase volume
and the gas density $V\rho=const$, that is true for normal "gas in a box" of
classical thermodynamics, is generally wrong for magnetized plasma.
An exception is the case of isotropic gas compression with $B \propto l^{-2}
\propto \rho ^{2/3}$ -- in this case the magnetic field acts as the walls of a
box, and this leads to the usual equations of state $\epsilon\propto\rho^{5/3}$
for the non-relativistic case or $\epsilon\propto\rho^{4/3}$ for relativistic
one. Here we use the proportionality of the particle energy to its perpendicular
momentum, that is true for two opposite cases -- either for  $p_{||}
\propto p_t$ (effective isotropization) or for $p_{||}\ll p_t$ (collisionless
case, only perpendicular momentum increases).

In the case of the accretion flow, however, the gas compression is
significantly anisotropic (the matter element is even stretched in the radial
 direction proportional to $r^{-1/2}$), and the magnetic field itself is not
perfectly frozen-in either. So, the relation of the phase volume element
spatial part to the density becomes
\begin{equation}
\frac{dV}{V}=-3\frac{dp_{t}}{p_t}=-\frac{3}{2}\frac{dB}{B}=-\frac{15}{8}
\frac{dr}{r}=-\frac54\frac{d\rho}{\rho}=-\frac54\frac{dn}{n}\mbox{   ,}
\label{eqn_phase_volume}
\end{equation}
where the radial dependencies
(\ref{eqn_radial_rho}),(\ref{eqn_radial_B}) 
of the accretion flow parameters are used.

Note that in contrary to assumptions made in most articles on this subject
(Bisnovatyi-Kogan \& Ruzmaikin \cite{bisnovatyi_1974}; Shapiro \cite{shapiro_1973b}; Meszaros \cite{meszaros};
Ipser \& Price \cite{ipser_1977,ipser_1982}; Mahadevan \& Quataert \cite{mahadevan}) we cannot use the first
law of thermodynamics in the form of "energy per particle"
\begin{equation}
d\left(\frac{\epsilon}{n}\right) = -pd\left(\frac1n\right) + ...
\label{eqn_energy_incorrect}
\end{equation}
(the dots here represent the contribution of non-adiabatic processes)
to describe the relation of the particle mean energy and the gas density as
the density changes don't reflect the behaviour of the particle "walls".
A correct form of this equation must take into account the fact that
the number of particles $N$ inside the spatial part of each particle conserving
phase volume element is no longer constant
\begin{equation}
d(\epsilon V) = -p dV + \frac{\epsilon}{n}dN + ...
\end{equation}
Using (\ref{eqn_phase_volume}), the change of the number of particles may be
expressed as
\begin{equation}
dN=nV\left(\frac{dV}{V}-\frac{d(1/n)}{(1/n)}\right)=-\frac14 V dn\mbox{ .}
\end{equation}
For relativistic gas the equation of state has the form $p=\epsilon/3$,
and so the single particle energy evolution is described by
\begin{equation}
d\left(\frac{\epsilon}{n}\right) = -\frac{5}{12} \epsilon d\left(\frac1n
\right) + ... \mbox{  ,}
\label{eqn_adiabatic_rel}
\end{equation}
where dots represent the possible contribution of non-adiabatic processes per
particle (thus, such particles behave like a gas with specific heat ratio $11/6$).

For non-relativistic gas in a similar manner ($p=\frac23\epsilon$)
\begin{equation}
d\left(\frac{\epsilon}{n}\right) = -\frac56 \epsilon d\left(\frac1n\right) +
... \mbox{  ,}
\label{eqn_adiabatic_nonrel}
\end{equation}
like the gas with a specific heat ratio of $17/12$.

Equations (\ref{eqn_adiabatic_nonrel}) and (\ref{eqn_adiabatic_rel})
may be rewritten in a form useful for comparison with the incorrect
expression (\ref{eqn_energy_incorrect}):
\begin{equation}
d\left(\frac{\epsilon}{n}\right) = -\frac54 pd\left(\frac1n\right) + ...
\mbox{  .}
\end{equation}

It is easily seen that correct consideration of adiabatic heating makes it
25\% more effective than in the case of ideal non-magnetized gas accretion
of Bondi (\cite{bondi_1952}), that has a large influence on luminosity and spectral
shape of the accretion flow emission.

\subsection{Radial temperature distribution}

Note that adiabatic heating alone does not change the shape of the particle
momentum distribution, so initially thermal distribution always stays thermal.

Current sheet spatial scales are usually much smaller than the whole
accretion flow one and, therefore, the fraction of the accelerated particles
is small, and so the total (significantly nonthermal) electron distribution
may be considered as a superposition of the purely thermal one for the background
flow particles and purely nonthermal for the ones accelerated in the
current sheets (this is the approach known as "hybrid plasma" of
Coppi (\cite{coppi})). Also we assume that for low accretion rates 
the non-adiabatic heating and radiative energy losses do not
change the shape of the thermal component distribution (while changing its mean
energy)
\begin{equation}
f(R,\gamma) = f_t(R,\gamma)+\zeta f_{nt}(R,\gamma).
\label{eqn_distribution_hybrid}
\end{equation}
Note that this distribution is not normalized, and only its shape has physical
meaning. So, for example, the ratio of nonthermal to thermal electron
densities at some radius $R$ may be expressed as
\begin{equation}
\frac{n_{nt}(R)}{n_{t}(R)} = \frac{\zeta f_{nt}(R)}{f_{t}(R)} \mbox{ ,}
\label{eqn_ratio_nonthermal_thermal}
\end{equation}
where $f_{nt}(R)$ and $f_{t}(R)$ are integrals of the corresponding
distribution functions over the range of $\gamma$.

Thermal particle distribution function may be written as
\begin{equation}
f_t(R,\gamma)=\frac{\sqrt{R}}{2\tau}\left(\frac{\gamma}{\tau}\right)^2\exp
{\left(-\frac{\gamma}{\tau}\right)},
\label{eqn_distribution_thermal}
\end{equation}
where the usual dimensionless expression for temperature $\tau=kT/m_ec^2$ is
used. This gives a Maxwellian local energy distribution and radial density
slope ${\rho\propto R^{-3/2}}$.

The temperature distribution $\tau(R)$ may be determined by solving the
energy balance equation taking into account heating due to adiabatic
compression and magnetic field dissipation and radiative losses.
Note that electrons become relativistic at some large radius $R_{rel}$,
while protons remain non-relativistic until the event horizon (Bisnovatyi-Kogan \&
Ruzmaikin \cite{bisnovatyi_1974}), and so their heating rates differ by a factor of $2$
(\ref{eqn_adiabatic_rel},\ref{eqn_adiabatic_nonrel}). In the case of
ineffective energy exchange (at low accretion rates collisions are very rare;
other mechanisms noted in Mahadevan \& Quataert (\cite{mahadevan}) are also ineffective);
this could have led to electrons much colder than protons, but due to
preferred heating of electrons by non-adiabatic processes (Bisnovatyi-Kogan \&
Lovelace \cite{bisnovatyi_2000, bisnovatyi_2001}) its energies always remain roughly of the same order of
magnitude. Of course their gamma-factors will differ by about $40$ times, and
so the main contribution to the accretion flow radiation is due to electrons.
Therefore, only electron temperature is interesting for us.

As it has been noticed before (\ref{eqn_dissipation_rate}), the
non-adiabatic heating rate may be expressed as
\begin{equation}
\Phi = (1-\xi)\frac{dE}{dVdt} = (1-\xi)\frac32\frac{v}{r}\frac{B^2}{8\pi}\mbox{ .}
\end{equation}

The main mechanism of radiative losses at low accretion rates is synchrotron
radiation. Its rate may be written as (Lightman \& Rybicki \cite{lightman})
\begin{equation}
\Lambda_{sync} = \frac43\sigma_Tc\overline{\gamma^2}\frac{B^2}{8\pi}n .
\label{eqn_lambda_sync}
\end{equation}

For a Maxwellian distribution
\begin{equation}
\overline{\gamma^2}=12\left(\frac{kT}{m_ec^2}\right)^2=12\tau^{2}.
\end{equation}

Taking into account non-adiabatic terms, the energy balance equation per particle
(\ref{eqn_adiabatic_nonrel}) may be written for a non-relativistic region of
the flow ($R>R_{rel}$) as 
\begin{equation}
\frac{d}{dr}\frac{\epsilon}{n}=-\frac{5}{4}\frac{\epsilon}{nr}+\frac{1}{v}\frac{\Lambda-\Phi}{n}\mbox{ .}
\end{equation}

For a non-relativistic electron gas $\epsilon=\frac32nkT$; also we may neglect
energy losses and rewrite it in variables $R$ and  $\tau$ as
\begin{equation}
\frac{d\tau}{dR}=-\frac{5}{4}\frac{\tau}{R}-(1-\xi)\frac{\alpha^2}{2}
\frac{m_p}{m_e}R^{-2} .
\label{temp_eqn_nonrel}
\end{equation}

This gives the value of the radius for $\tau=1$, $R_{rel}\approx6400$
for $\xi=0.1$, that significantly exceeds $R_{rel}\approx1300$ in the ideal
gas approximation used by Bisnovatyi-Kogan \& Ruzmaikin (\cite{bisnovatyi_1974}).

In deeper regions, electrons become relativistic, their energy density is
$\epsilon=3p=3nkT$, and so for $R<R_{rel}$ the energy balance equation
(\ref{eqn_adiabatic_rel})  combined with (\ref{eqn_radial_rho}),
(\ref{eqn_radial_B}) and (\ref{eqn_dotm}) gives
\begin{equation}
\frac{d\tau}{dR}=-\frac58\frac{\tau}{R}-(1-\xi)\frac{\alpha^2}{4}\frac{m_p}{
m_e}R^{-2}+\frac43\frac{m_p}{m_e}\frac{\dot m \tau^2}{R^2}.
\label{temp_eqn}
\end{equation}

The boundary condition is ${\tau(R_{rel})=1}$.

An analytical solution of this equation in general is very difficult,
but for low accretion rates we may neglect the influence of radiative
losses and get a solution in the form
\begin{equation}
\tau(R)=(1-\xi)\frac{2\alpha^2}{3}\frac{m_p}{m_e}R^{-1}+\left(1-(1-\xi)
\frac{2\alpha^2}{3R_{rel}}\frac{m_p}{m_e}\right)\left(\frac{R_{rel}}{R}
\right)^{5/8}.
\end{equation}
This expression may be substituted into (\ref{eqn_distribution_thermal}) to
get the final expression for the thermal electron distribution.

The applicability of this approximation may be estimated by comparing the
timescales of the electron radiative energy losses and of the free-fall near
the event horizon
\begin{equation}
\frac{t_{ff}}{t_{rad}}=\frac{\alpha}{3}\frac{m_p}{m_e}\gamma\dot m <1 \mbox{\ \ \ \ for\ \ \ \ } \dot m< 10^{-5} ,
\end{equation}
where $\gamma\approx100$ is assumed. For higher accretion rates equation (\ref{temp_eqn})
may be easily solved numerically.

\subsection{Distribution function of the non-thermal component}

We can build an expression for nonthermal component distribution function.
We may assume that all electrons are relativistic as only
these have significant observational appearances, and electrons are
relativistic in the most interesting regions near the horizon (well inside the
relativization radius $R_{rel}\approx6000$). So, the fraction of
low-energy nonthermal particles is negligible, and nearly all electrons have
a "lifetime" not less than the characteristic free-fall time scale
(see Sec. \ref{sec_acceleration}).

So, we may neglect nonthermal particle interactions with electrons and
plasmons of the background flow (see discussion in Sec. \ref{sec_acceleration})
and assume that they evolve due to adiabatic heating and synchrotron energy
losses only. Note that this leads to the situation where thermal electron
energy grows slightly faster than the nonthermal one due to additional heating by
plasma oscillations ejected by current sheets.

The energy evolution of single nonthermal electron is described by
\begin{equation}
\frac{d\gamma}{dR} = \frac13\frac{m_p}{m_e}\dot m\frac{\gamma^2}{R^2} -
\frac58\frac{\gamma}{R},
\label{eqn_gamma_evolution}
\end{equation}
where the first term corresponds to synchrotron losses and the second one 
to adiabatic heating.

For the initial energy $\gamma_0$ at $R_0$ it has the solution
\begin{equation}
\gamma = \frac{\gamma_0}{C_1(R,R_0)\gamma_0+C_2(R,R_0)} \\
\label{eqn_gamma_evolution_solution}
\end{equation}
where
\begin{eqnarray}
\nonumber C_1(R,R_0) & = & \frac{A}{R}\left(1-\left[\frac{R}{R_0}\right]^{13/8}
\right) \\
\nonumber A & = & \frac{8}{39}\frac{m_p}{m_e}\dot m  \\
\nonumber C_2(R,R_0) & = & \left(\frac{R}{R_0}\right)^{5/8}.
\end{eqnarray}

The nonthermal component at some radius $R$ consists of non-interacting
electron beams generated at all radii $R_0>R$. The evolution of the distribution
function of each such beam (assuming that the ejection process is stationary and
the initial beam distribution function is $f_b(R_0,\gamma_0)$) is as follows:
\begin{eqnarray}
\nonumber f_b(R,\gamma) & = & f_b(R_0,\gamma_0)\frac{d\gamma_0}{d\gamma} \\
& = & f(R_0,\gamma_0)\frac{(C_1(R,R_0)\gamma_0+C_2(R,R_0))^2}{C_2(R, R_0)}.
\label{eqn_distribution_f_single}
\end{eqnarray}

The initial beam distribution has a form (see (\ref{eqn_distribution_ejected}))
\begin{equation}
f_b(R_0,\gamma_0) = \frac{f_b(R_0)}{\Gamma(R_0)}\left(\frac{\Gamma(R_0)}
{\gamma_0}\right)^3\exp{\left(-\frac{\Gamma(R_0)}{\gamma_0}\right)},
\label{eqn_distribution_f0}
\end{equation}
where the mean energy
$$\overline{\gamma_0} = \Gamma = \Delta\tau$$
is assumed to be by a fixed factor $\Delta$ greater than the local thermal
component one and $f_b(R)$ describes the radial distribution of the accelerated
particle ejection rate $dN/dt$. The latter may be computed using
(\ref{eqn_dissipation_radial}) as
\begin{equation}
\frac{dN}{dR dt} = \frac{1}{m_e c^2 \Gamma(R)} \frac{\xi dE}{dR dt} 
= \frac{6\pi\alpha^2}{4}\frac{m_p}{m_e}\frac{r_g c}{\sigma_T}\frac{\xi
\dot m}{R^2\Gamma(R)},
\label{eqn_ejection_radial}
\end{equation}
which gives for the total ejection rate
\begin{equation}
\frac{dN}{dt} = \frac{6\pi\alpha^2}{4}\frac{m_p}{m_e}\frac{r_g c}{\sigma_T}
\xi \dot m \int\limits_1^{\infty}\frac{dR}{R^2\Gamma(R)},
\label{eqn_ejection_total}
\end{equation}
and the radial distribution of particle ejection
\begin{equation}
f_b(R_0)= \frac{a_0}{\Gamma(R_0) R_0^2} \mbox{ , where } 
a_0 = \left(\int\limits_1^{\infty}\frac{dR_0}{\Gamma(R_0) R_0^2}\right)^{-1}.
\label{eqn_beam_radial}
\end{equation}

By integrating (\ref{eqn_distribution_f_single}) over all $R_0 > R$ we may
get final expression for nonthermal electron distribution which consists of
all beam electrons ejected at greater distances
\begin{eqnarray}
\nonumber
f_{nt}(R, \gamma) & = & \int\limits_R^{\infty}\frac{a_0 dR_0}{R_0^2}
\frac{\Gamma(R_0)}{\gamma_0^3}
  \frac{\left[C_1(R, R_0)\gamma_0+C_2(R, R_0)\right]^2}{C_2(R, R_0)}\\
   & & \times \exp{\left(-\frac{\Gamma(R_0)}{\gamma_0}\right)}.
\label{eqn_distribution_f}
\end{eqnarray}

Now we determine the $\zeta$ coefficient of the total electron distribution
(\ref{eqn_distribution_hybrid}). By combining
(\ref{eqn_ratio_nonthermal_thermal}) with trivial expressions
for the energy density ratio
\begin{equation}
\frac{\epsilon_{nt}(R)}{\epsilon_t(R)} = \frac{n_{nt}(R)}{n_t(R)}
\frac{\overline \gamma_t(R)}{\overline \gamma_{nt}(R)}
\end{equation}
and for the energy density of thermal electrons with a Maxwellian value $\overline
\gamma_t(R) = 3 \tau(R)$
\begin{equation}
\epsilon_t = 3 m_e c^2 \tau(R) n_t
\end{equation}
we may get
\begin{equation}
\zeta = \frac{\epsilon_{nt}(R)\sqrt{R}}{m_e c^2 n_t(R)}\left(\int\limits_1^
{\infty}\gamma f_{nt}(R, \gamma) d\gamma \right)^{-1}.
\label{eqn_zeta_epsilon_nonthermal}
\end{equation}

The fraction of nonthermal particles is always small, so we may replace here
the thermal particle number density $n_t(r)$ with the total one $n(R)$.

The nonthermal electron energy density at $R$ may be computed by considering
an elementary spherical shell ${R \div R+\delta R}$ and integrating the nonthermal
energy release in it during its whole backtrace free-fall history
\begin{equation}
\epsilon_{nt}(R) = \frac1{4\pi r_g^3 R^2 \delta R} \int\limits_R^{\infty}
\frac{\delta R_0 dR_0}{-\frac{dR_0}{dt}}
\left(\frac{\xi dE}{dR_0 dt}\right) \frac{\overline \gamma_b (R)}
{\Gamma(R_0)},
\label{eqn_epsilon_nonthermal}
\end{equation}
where the latter multiplicative term corresponds to amplification of the
mean gamma-factor of each accelerated electron beam ejected at $R_0$ in free-fall until $R$. This quantity may
be written using the results of Appendix \ref{appendix_beam_moments} as
\begin{equation}
\frac{\overline \gamma_b (R)}{\Gamma(R_0)} = \Fcurve_1\left(\left
(\frac{R}{R_0}\right)^{5/8},\frac{A\Gamma(R_0)}{R}\left(1-\left
[\frac{R}{R_0}\right]^{13/8}\right)\right).
\end{equation}

This combined with the expression for the local nonthermal energy
dissipation rate (\ref{eqn_dissipation_radial}) and with the scaling for the
given spherical shell thickness at free-fall ${\delta R_0 \propto R_0^{-1/2}}$
leads to the final expression for the $\zeta$ coefficient

\begin{eqnarray}
\nonumber 
\zeta & = & \frac34\frac{m_p}{m_e} \xi\alpha^2\sqrt{R} \left(\int\limits_1^
{\infty}\gamma f_{nt}(R, \gamma) d\gamma \right)^{-1}\\
 & & \times \int\limits_R^{\infty} \frac{dR_0}{R_0^2} \Fcurve_1\left(\left
(\frac{R}{R_0}\right)^{5/8},\frac{A\Gamma(R_0)}{R}\left(1-\left[\frac{R}
{R_0}\right]^{13/8}\right)\right).
\label{eqn_zeta_final}
\end{eqnarray}

\begin{figure}
{\centering \resizebox*{1\columnwidth}{!}{\includegraphics[angle=270]
{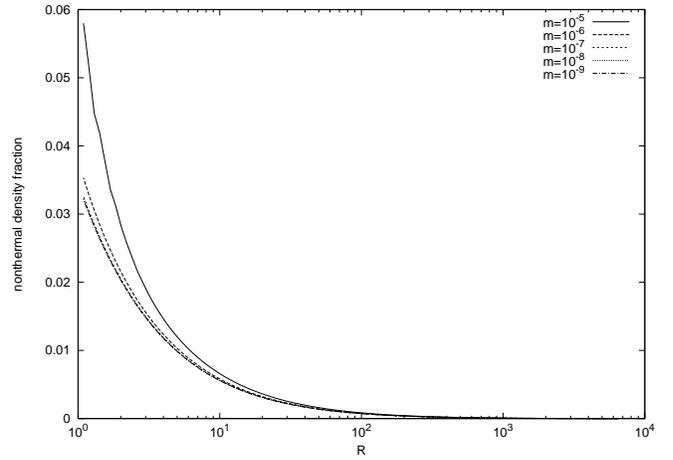}} \par}
\caption{Nonthermal electron density fraction $n_{nt}(R)/n(R)$ for an $M=10M_{\odot}$ black hole,
$\xi=0.1$ and the number of accretion rates.}
\label{fig_fraction}
\end{figure}

\begin{figure}
{\centering \resizebox*{1\columnwidth}{!}{\includegraphics[angle=270]
{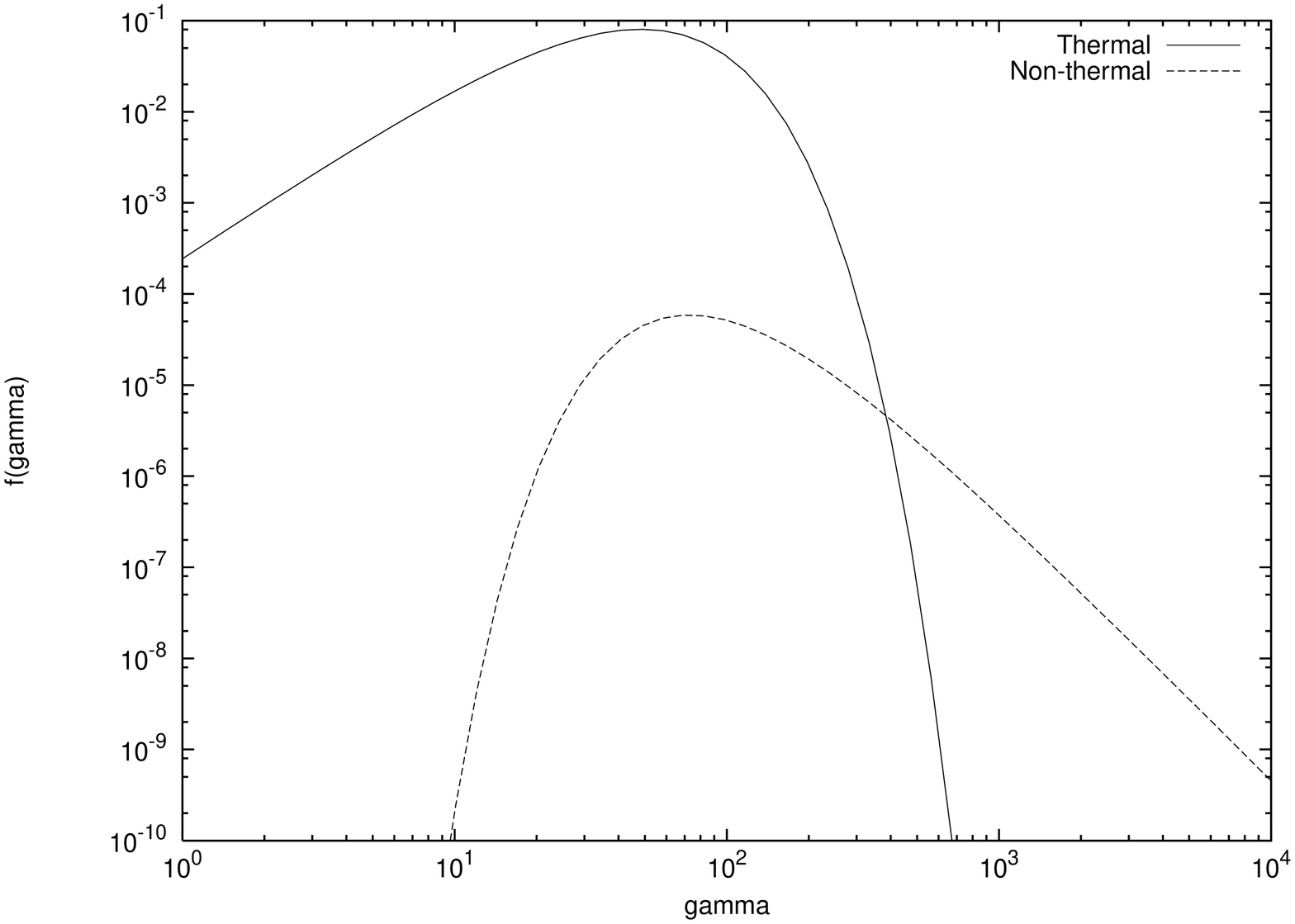}} \par}
{\centering \resizebox*{1\columnwidth}{!}{\includegraphics[angle=270]
{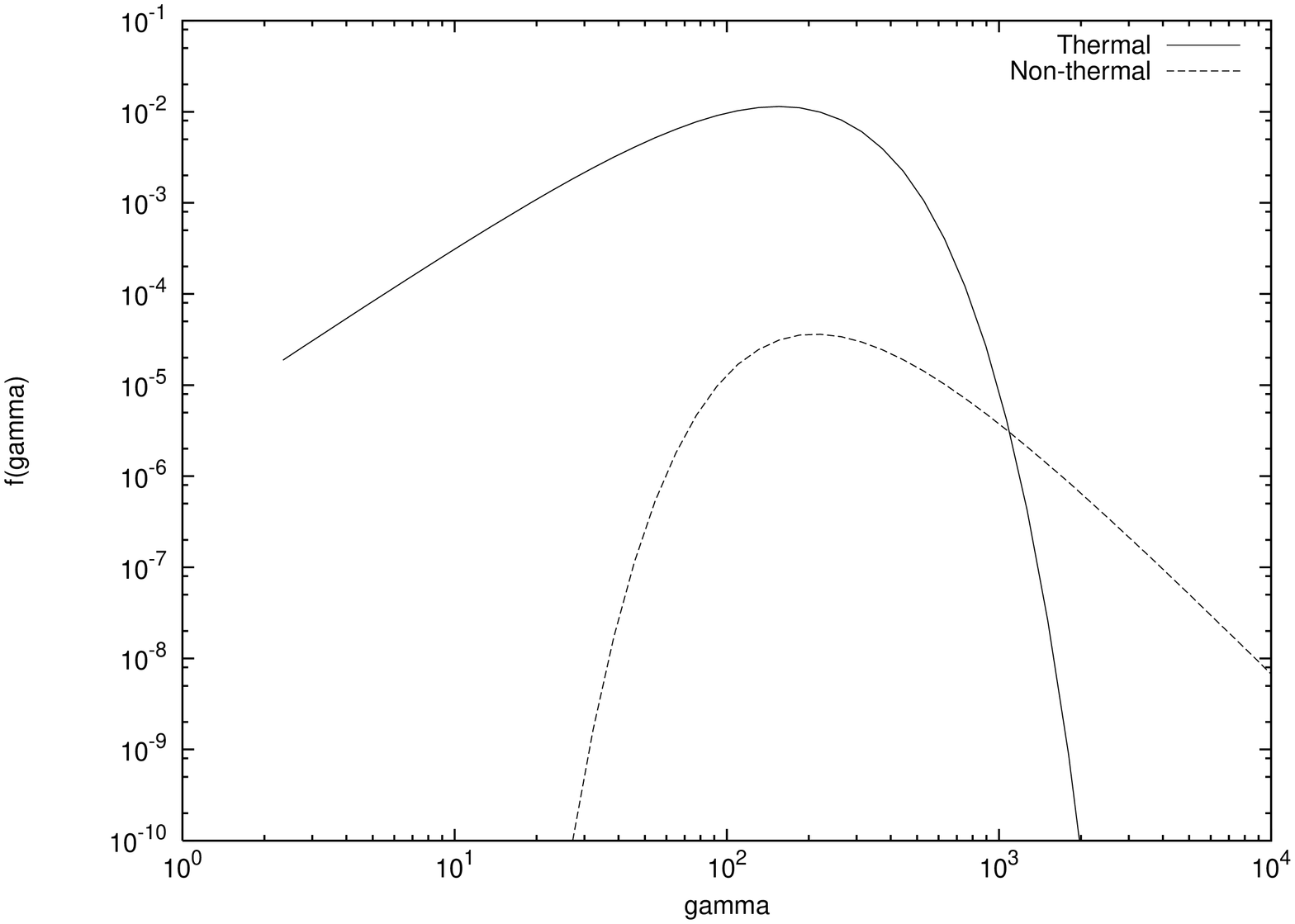}} \par}
\caption{Energetic distributions of thermal $f_t(R, \gamma)$ and nonthermal $f_{nt}(R, \gamma)$ electron components
for distances of $R = 50r_g$ (first panel) and $R = 10r_g$ (second panel), respectively, for $\Delta=10$.}
\label{fig_distribution}
\end{figure}

A sample fraction of the nonthermal to total densities for different
accretion rates is shown in Fig.\ref{fig_fraction} as a function of distance
from the black hole. This fraction is small. Fig.\ref{fig_distribution}
shows the shapes of thermal and non-thermal distributions for two different
distances at low accretion rates.

\section{Emission spectrum}
\label{sec_spectra}

The emission spectrum of a single electron for an observer at infinity has
the shape (Lightman \& Rybicki \cite{lightman}; Shapiro \cite{shapiro_1973a})
\begin{equation}
L_{\nu}=2\pi\int\limits_{-1}^{\cos{\theta^*}}j_{\nu'}\frac{1-\beta^2}
{(1-\beta\cos{\theta})^2}d\cos{\theta} ,
\end{equation}
where relativistic effects of time contraction, gravitational redshift,
Doppler effect and the capture of some emission fraction by the event horizon
are taken into account. The event horizon angular size for a free-falling
emitter is
$$|\cos{\theta^*}|=\sqrt{1-\frac{27}{4R^2}(1-\frac1R)},$$
where ${\cos{\theta^*<0}}$ for ${R<1.5}$. The quantity
$$\beta=\frac{dr}{dt}\frac{1}{1-r_g/r}=\frac{v/c}{\left(v^2/c^2+1-r_g/r
\right)^{1/2}}$$
represents the falling velocity of matter in the distant observer frame,
and the frequency shift is given by
$$\nu'=\nu\frac{1-(v/c)\cos{\theta}}{\sqrt{(1-v^2/c^2)(1-1/R)}}.$$

By substituting synchrotron emissivity from Lightman \& Rybicki (\cite{lightman})
we get
\begin{equation}
j_{\nu}=\frac{\sqrt{3}e^3 B\sin{\psi}}{4\pi m_ec^2}F\left(\frac{\nu'}
{\nu_c}\right),
\end{equation}
\begin{equation}
F(x)=x\int\limits_x^{\infty}K_{5/3}(\xi)d\xi ,
\end{equation}
\begin{equation}
\nu_c=\frac{3\gamma^2eB\sin{\psi}}{4\pi m_ec.}.
\end{equation}

Taking into account the fact that adiabatic compression increases
the perpendicular electron momentum only (and so we may take $\sin{\psi}\approx1$
for the pitch-angle) and convolving with the electron distribution
(\ref{eqn_distribution_hybrid}), the final expression for the accretion flow
emission spectrum is
\begin{equation}
L_{\nu}\propto\int\limits_1^{\infty}R^{-5/4}dR\int\limits_{1}^{\infty}f
(R,\gamma)\int\limits_{-1}^{\cos{\theta^*}}\frac{1-\beta^2}{1-\beta\cos
{\theta}}F\left(\frac{\nu'}{\nu_c}\right)d\cos{\theta}d\gamma .
\label{eqn_spectrum_shape}
\end{equation}

\begin{figure}
\center
{\centering \resizebox*{1\columnwidth}{!}{\includegraphics[angle=270]{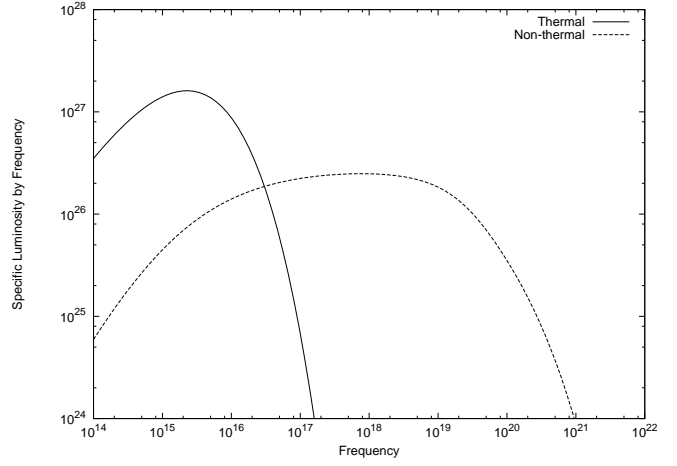}} \par}
\caption{Decomposition of a single black hole (with the mass $10 M_{\odot}$)
emission spectrum into thermal and nonthermal parts. The accretion rate is
$1.4\cdot 10^{10}$ g/s, which corresponds to $\dot m=10^{-8}$, and $\xi=0.1$.}
\label{fig_spectrum_decomposed}
\end{figure}

\begin{figure}
\center
{\centering \resizebox*{1\columnwidth}{!}{\includegraphics[angle=270]{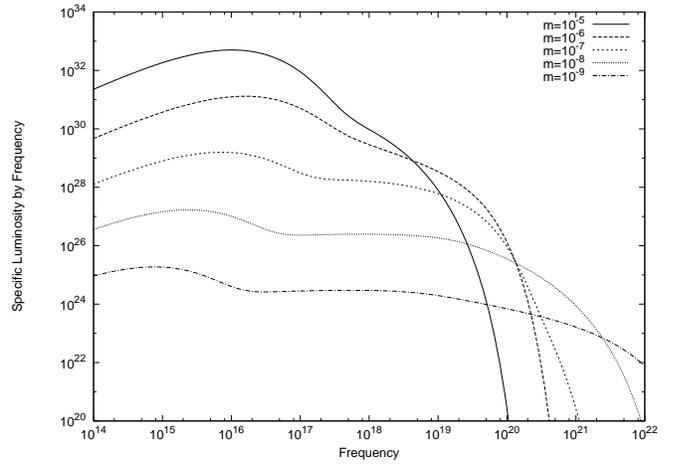}} \par}
\caption{Accreting $10M_{\odot}$ black hole spectra for accretion rates from
$\dot m=10^{-9}$ till $\dot m=10^{-5}$ ($\dot M=1.4\cdot 10^{9}$ g/s and
$\dot M=1.4\cdot 10^{13}$ g/s, correspondingly), for $\xi=0.1$. }
\label{fig_spectrum_multi}
\end{figure}

Fig. \ref{fig_spectrum_decomposed} shows a decomposition of the accretion
flow spectrum into thermal and nonthermal parts. The fraction of the
nonthermal emission is small in the optical range, but dominates in harder spectral
bands. The spectrum becomes flatter with the accretion rate decrease due to
the increase of the electron radiative loss timescale. The high-energy spectral
cut-off is determined by radiative energy losses and by the upper limits of
the accelerated electron gamma-factor (see Fig.\ref{fig_spectrum_multi}).

\section{Luminosity}

The luminosity of the accretion flow may be computed by integrating
expression (\ref{eqn_spectrum_shape}) (Shapiro \cite{shapiro_1973a}; Ipser \& Price \cite{ipser_1977})
\begin{eqnarray}
\nonumber
L & = & 8\pi^2\int\limits_{1}^{\infty}R^2dR\int\limits_{-1}^{\cos{\theta}^*}
\int\limits_{0}^{\infty}j_{\nu'}d\nu'
\\
& & \times {\left(1-\frac{r_g}{r}\right)^{1/2}\left(1-\beta^2\right)^{3/2}}
(1-\beta\cos{\theta})^{-3} d\cos{\theta}.
\label{eqn_luminosity}
\end{eqnarray}

Dividing by $\dot Mc^2$ and taking into account the expression for synchrotron
luminosity of a single electron (\ref{eqn_lambda_sync}), we get the
expression for the efficiency of the thermal component emission
\begin{equation}
\eta_t=4\dot m\int\limits_1^{\infty}\frac{\tau^2}{R^2}K(R)dR ,
\label{eqn_efficiency_thermal}
\end{equation}
where 
\begin{equation}
K(R)=\frac12\int\limits_{-1}^{\mu^*}\frac{(1-1/R)^{1/2}(1-\beta^2)^{3/2}}
{(1-\beta x)^3}dx .
\end{equation}

At low accretion rates ($\dot m \ll 10^{-5} $) a reasonably good
approximation for this quantity is
\begin{equation}
\eta_t=4.5\cdot10^{4}\dot m,
\end{equation}
which corresponds to the thermal luminosity of
\begin{equation}
L_t = 9.6\cdot10^{33} M_{10}^3 n_1^2 (V^2 + c_s^2)_{16}^{-3} \mbox{  erg/s}.
\end{equation}

The efficiency of the nonthermal component emission may be easily estimated
from the radial distribution of the dissipated magnetic energy $dE/dRdt$,
its fraction $\xi$ carried out by accelerated particles and the evolution of
its mean square of the gamma-factor during the fall towards the horizon
$\overline \gamma_b^2(R)$ as
\begin{equation}
\eta_{nt}=\frac{\xi}{4}\int\limits_1^{\infty}\frac{a_0 dR_0}{\Gamma^2
(R_0)R_0^2}\int\limits_1^{R_0}\left(\frac{d\overline\gamma_b}{dR}
\right)_{emis}K(R)dR ,
\end{equation}
where the mean radiative energy losses of a single electron are
\begin{equation}
\left(\frac{d\overline\gamma_b}{dR}\right)_{emis}=\frac13\frac{m_p}{m_e}
\dot m\frac{\overline{\gamma_b^2}}{R^2} .
\end{equation}
The evolution of the mean square of gamma-factor $\overline \gamma_b^2(R)$
of the beam is given by expression (\ref{eqn_mean_gamma2}) derived in
Appendix \ref{appendix_beam_moments}

So, for the efficiency of the non-thermal component emission we have
\begin{eqnarray}
\nonumber 
\eta_{nt} & = & \frac{a_0 \xi \dot m}{12}\int\limits_1^{\infty}\frac{dR_0}{R_0^2}
\int\limits_1^{R_0}\frac{dR}{R^2}K(R)
\\ 
& & \times \Fcurve_2\left(\left(\frac{R}{R_0}\right)^{5/8},\frac{A\Gamma(R_0)}{R}
\left(1-\left[\frac{R}{R_0}\right]^{13/8}\right)\right).
\end{eqnarray}

\begin{figure}
\center
{\centering \resizebox*{1\columnwidth}{!}{\includegraphics[angle=270]
{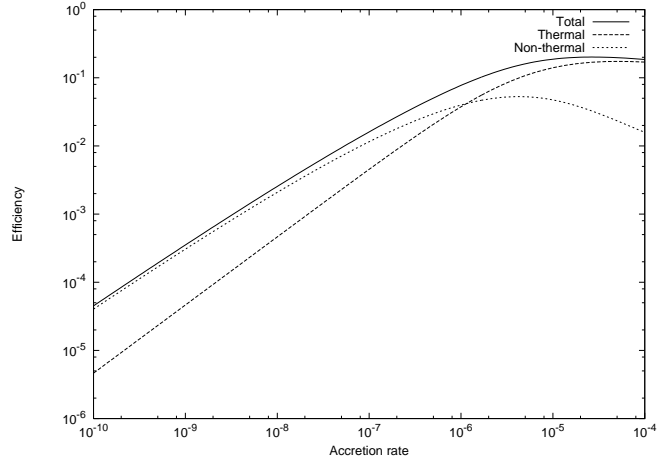}} \par}
\caption{Efficiencies of the synchrotron emission of thermal and non-thermal
electron components of the accretion flow.}
\label{fig_efficiency}
\end{figure}

The results of numerical computations according to these formulae may be
seen in Fig.\ref{fig_efficiency}.

\begin{table*}
\caption{Thermal and nonthermal luminosity, magnitudes, fluxes and flares rate
of a 10 M$_{\odot}$ black hole in a 
various ISM conditions at a 100 pc distance.
}
\label{table_luminosity}
\[
\begin{array}{ccccccccccccc}
\hline
\noalign{\smallskip}
n & V  & \dot m & L_t & L_{nt} & m_V & m_V^{nt} & \rm{EPIC}^1 & \rm{HRC}^2 & \rm{JEM-X}^3 & \rm{IBIS}^4 & \rm{XRT}^5 & \rm{Flare\ rate}^6\\
\rm{[cm^{-3}]} & \rm{[km/s]} &  & \rm{[erg/s]} & \rm{[erg/s]} & & & \rm{[erg/s/cm^2]} & \rm{[erg/s/cm^2]} & \rm{[erg/s/cm^2]} & \rm{[erg/s/cm^2]} & \rm{[erg/s/cm^2]} & \rm{[10^3/s]} \\
\noalign{\smallskip}
\hline
\noalign{\smallskip}
\noalign{\mbox{Hot clouds, T$\approx10^6$K, $\xi$=0.1}}
\noalign{\smallskip}
\hline
\noalign{\smallskip}
2\cdot10^{-3} & 10 & 6.5\cdot 10^{-11} & 2.5 \cdot 10^{23} & 2.3 \cdot 10^{24} & 36.1 & 38.7 & 5.5\cdot10^{-20} & 5.2\cdot10^{-20} & 3.8\cdot10^{-20} & 7.6\cdot10^{-20} & 5.2\cdot10^{-20} & 7.5 \\
2\cdot10^{-3} & 50 & 5.1 \cdot 10^{-11} & 1.5 \cdot 10^{23} & 1.4 \cdot 10^{24} & 36.6 &  39.1 & 3.6\cdot10^{-20} & 3.4\cdot10^{-20} & 2.5\cdot10^{-20} & 74.9\cdot10^{-20} & 3.3\cdot10^{-20} & 7.4 \\
2\cdot10^{-3} & 100 & 2.9 \cdot 10^{-11} & 4.9 \cdot 10^{22} & 4.9 \cdot 10^{23} & 37.9 & 40.1 & 1.3\cdot10^{-20} & 1.2\cdot10^{-20} & 9.1\cdot10^{-21} & 1.7\cdot10^{-20} & 1.2\cdot10^{-20} & 7.5 \\
\noalign{\smallskip}
\hline
\noalign{\smallskip}
\noalign{\mbox{Warm clouds, T$\approx10^4$K, $\xi$=0.1}}
\noalign{\smallskip}
\hline
\noalign{\smallskip}
0.2 & 10 & 2.9\cdot 10^{-6} & 2.9 \cdot 10^{32} & 1.9 \cdot 10^{32} & 15.6 & 20.0 & 2.2\cdot10^{-11} & 2.3\cdot10^{-11} & 7.2\cdot10^{-13} & 1.0\cdot10^{-13} & 2.2\cdot10^{-11} & 10 \\
0.2 & 50 & 7.9 \cdot 10^{-8} & 3.5 \cdot 10^{29} & 9.7 \cdot 10^{29} & 22.1 & 26.4 & 3.6\cdot10^{-14} & 3.5\cdot10^{-14} & 1.8\cdot10^{-14} & 1.2\cdot10^{-14} & 3.5\cdot10^{-14} & 7.7 \\
0.2 & 100 & 1.0 \cdot 10^{-8} & 6.3 \cdot 10^{27} & 2.8 \cdot 10^{28} & 25.8 & 29.8 & 7.7\cdot10^{-16} & 7.3\cdot10^{-16} & 5.3\cdot10^{-16} & 7.2\cdot10^{-16} & 7.3\cdot10^{-16} & 7.6 \\
\noalign{\smallskip}
\hline
\noalign{\smallskip}
\noalign{\mbox{Warm clouds, T$\approx10^4$K, $\xi$=0.5}}
\noalign{\smallskip}
\hline
\noalign{\smallskip}
0.2 & 10 & 2.9\cdot 10^{-6} & 1.5 \cdot 10^{32} & 7.6 \cdot 10^{32} & 15.7 & 18 & 1.8\cdot10^{-11} & 1.8\cdot10^{-11} & 2.5\cdot10^{-12} & 3.9\cdot10^{-13} & 1.8\cdot10^{-11} & 31 \\
0.2 & 50 & 7.9 \cdot 10^{-8} & 1.6 \cdot 10^{29} & 3.3 \cdot 10^{30} & 22.2 & 24.3 & 1.2\cdot10^{-13} & 1.2\cdot10^{-13} & 6.8\cdot10^{-14} & 4.3\cdot10^{-14} & 1.2\cdot10^{-14} & 24 \\
0.2 & 100 & 1.0 \cdot 10^{-8} & 2.9 \cdot 10^{27} & 9.4 \cdot 10^{28} & 26 & 27.8 & 3.0\cdot10^{-15} & 2.8\cdot10^{-15} & 2\cdot10^{-15} & 2.8\cdot10^{-15} & 2.8\cdot10^{-15} &  23 \\
\noalign{\smallskip}
\hline
\noalign{\smallskip}
\noalign{$^1$ XMM EPIC band flux. Sensitivity is $10^{-14}$ erg/s$\cdot$cm$^2$ for typical survey exposures.}
\noalign{\smallskip}
\noalign{$^2$ Chandra HRC band flux. Sensitivity ranges from $10^{-14}$ erg/s$\cdot$cm$^2$ to $10^{-16}$ erg/s$\cdot$cm$^2$ (deep surveys)}
\noalign{\smallskip}
\noalign{$^3$ INTEGRAL Jem-X band flux. Sensitivity is $4\cdot10^{-11}$ erg/s$\cdot$cm$^2$ for 100 ksec exposure}
\noalign{\smallskip}
\noalign{$^3$ INTEGRAL IBIS band flux. Sensitivity is $10^{-15}$ erg/s$\cdot$cm$^2$ for 1000 ksec exposure}
\noalign{\smallskip}
\noalign{$^5$ Swift XRT band flux. Sensitivity is $5\cdot10^{-13}$ erg/s$\cdot$cm$^2$ for 1 ksec exposure}
\noalign{\smallskip}
\noalign{$^6$ Flare rate is for the flares with mean luminosity greater than 0.1\% of mean nonthermal one}

\end{array}
\]
\end{table*}

\section{Properties of flares}
\label{sec_flares}

We now discuss some temporal properties of the accretion flow emission.
Neglecting the complex spatial structure of the accretion flow we may
consider the thermal electron component emission $L_0$ as a constant background
with a highly variable nonthermal emission component superimposed.
The variability of the latter is mostly due to the discrete nature of particle
acceleration in the current sheets. Assuming for simplicity that each
non-thermal electron accelerates only once and then free-falls and evolves
due to adiabatic heating and radiative losses only, we get for the light
curve of a single flare (i.e. a beam of $N$ electrons ejected by a
reconnection event at some distance $R_0$ with distribution function
(\ref{eqn_distribution_ejected}) and mean gamma-factor $\Gamma(R_0)=
\Delta\tau(R_0)$) an expression
\begin{equation}
\Delta L = \frac43 \sigma_T c \left( \overline \gamma_b^2 - 12\tau^2 \right)
\frac{B^2}{8\pi} N K(R) ,
\label{eqn_flare_lightcurve}
\end{equation}
where the quantity $\overline \gamma_b^2$ is derived in Appendix
\ref{appendix_beam_moments} and represents the mean square of the gamma-factor
of the electron beam ejected at $R_0$ and free-fallen to $R$. The mean square
of the Maxwellian distribution (\ref{eqn_distribution_thermal}) gamma-factor
value of $12\tau^2$ is also taken into account (we need to subtract this term
since each electron must occur in the total luminosity expression only once).
The $K(R)$ coefficient describes relativistic effects of the radiation
reduction.

By introducing an effective volume of the current sheet $V=N/n(R_0)$ and
dividing by the mean luminosity of the nonthermal component $L_t=\eta_{nt} \dot M
c^2$, we get
\begin{equation}
\frac{\Delta L}{L_{nt}} = \frac{\dot m}{12\pi \eta_{nt} R_0^{3/2}}\left(\frac{V}
{r_g^3}\right) \frac{\overline{\gamma_b^2}-12\tau^2}{R^{5/2}} K(R) .
\end{equation}
Transition to the light curve may be performed by substituting the dependence
of distance on time for the falling gas element as
\begin{equation}
R=\left(R_0^{3/2}-\frac{3\alpha}{2}\frac{c}{r_g}t\right)^{2/3}\mbox{ .}
\end{equation}

\begin{figure}
{\centering \resizebox*{1\columnwidth}{!}{\includegraphics[angle=0]{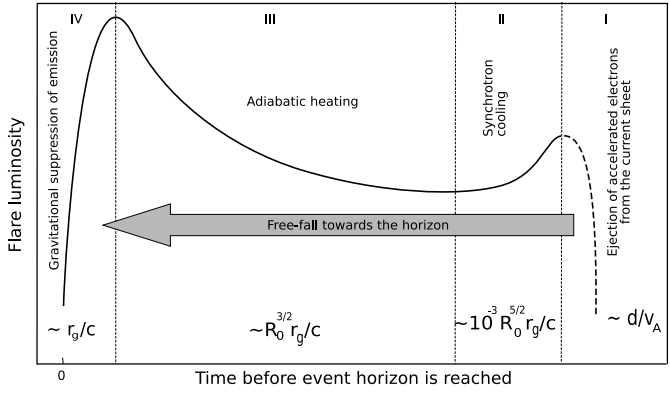}} \par}
\caption{Internal structure of a flare as a reflection of the electron cloud evolution. The prevailing
physical mechanisms defining the observed emission are denoted and typical durations of the
stages are shown.}
\label{fig_flare_structure}
\end{figure}

The temporal structure of the flare is complex (see
Fig.\ref{fig_flare_structure}). It consists of several parts with
the domination of different physical processes on the electron cloud
evolution. The first stage is the acceleration of the electrons in the
current sheet (see section \ref{sec_acceleration}), which occurs on the
Alfvenic time scale, but the exact temporal structure depends on
the properties of the current sheet and thus is beyond the
scope of the article. The second stage is the fast emission decay due to synchrotron
energy losses on the corresponding energy scale; it lasts until the
establishment of the equilibrium between energy losses and adiabatic
heating rates. The third stage is generally the longest one, at least
for $R_0 < 10^3$,  and has the time scale of a free-fall. Finally, near
the event horizon the relativistic effects (time contraction,
gravitational redshift, Doppler effect and the photons capture by the
BH) prevail and flare emission decays on the $r_g/c$ time scale as the
electron cloud approaches the horizon. This stage is very important as
it directly reflects the metrics near the black hole.

\begin{figure}
{\centering \resizebox*{1\columnwidth}{!}{\includegraphics[angle=270]
{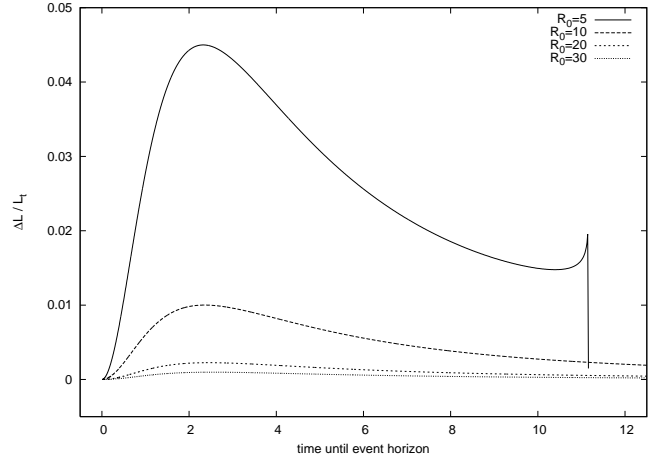}} \par}
\caption{Light curves of separate non-thermal flares -- beams of accelerated
electrons ejected at $R_0$ -- for $\dot m = 10^{-8}$. Time is measured in units of $r_g/c$,
luminosity in units of total thermal luminosity. The stage of particle
acceleration (which defines the flare front) is omitted as its shape depends 
on an unknown temporal structure of the particle acceleration process, which is beyond the
scope of our consideration.}
\label{fig_flare_lightcurve}
\end{figure}

Fig.\ref{fig_flare_lightcurve} shows the sample light curves of flares with
$R_0=5$, $R_0=10$, $R_0=20$ and $R_0=30$, for the accretion rate $\dot m=10^{-8}$ and the
equivalent volume $V=r_g^3$. The unknown first stage (flare front) is excluded.

\begin{figure}
{\centering \resizebox*{1\columnwidth}{!}{\includegraphics[angle=270]{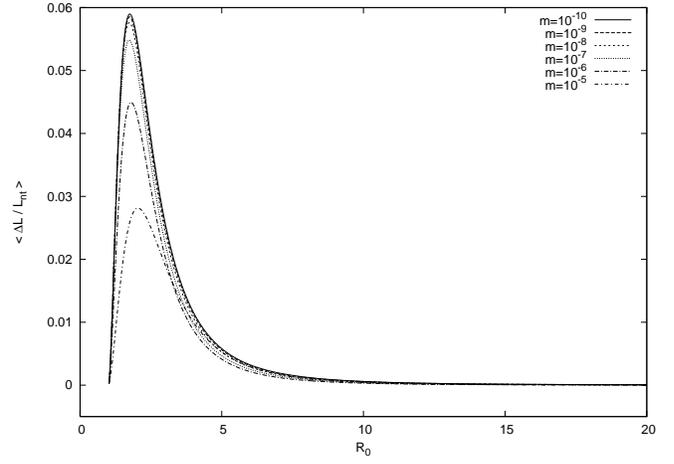}} \par}
\caption{Mean amplitude of a flare as a function of ejection radius $R_0$.}
\label{fig_flare_amplitude}
\end{figure}

By averaging over the flare light curve we may get the dependence of mean flare amplitude
$<\Delta L / L_{nt}>$ on the ejection radius $R_0$. This is shown in Fig.\ref{fig_flare_amplitude}.
It is clear that the strongest flares are produced by reconnections very near the event 
horizon (at several $r_g$).
The rate of a flares may be easily computed as 
\begin{equation}
\frac{d\aleph}{dR_0dt} = \frac{dE}{dR_0dt} \frac{\xi}{E \Gamma(R_0)}\int\limits_1^{R_0} \left(\frac{d\overline\gamma_b}{dR}
\right)_{emis}K(R)dR
\end{equation}
where $E$ is the total energy of a flare as is seen at infinity
\begin{equation}
E = L_{nt} \int <\Delta L / L_{nt}> dt
\end{equation}

By integrating this expression over the range of $R_0$ where $<\Delta L / L_{nt}>$ 
is greater than some threshold value we may get the rate of
flares with given amplitudes. Results of such computation for the rate
of flares with an amplitude greater than 0.1\% of the mean nonthermal
luminosity ($<\Delta L / L_{nt}> \ > 10^{-3}$) are given in the Table
\ref{table_luminosity}.

The fine temporal structure of these flares may be related to the motion of
the electron beam in the magnetic field (this motion, as it has been noted
before, is finite; electrons do not leave the small volume with the size
determined by the Larmor radius and magnetic field line topology). So, if the
characteristic size of the beam (the characteristic timescale of particle
ejection multiplied by the speed of light) is smaller than that of the
magnetic field loops, then we see the beam emission only when it is
pointed towards us. The timescale of such flares is much shorter than
the free-fall one, and they can be detected only in high time resolution observations.
On the other hand,
their properties reflect the magnetic field structure, and so its search is very important.

\begin{figure}
{\centering \resizebox*{1\columnwidth}{!}{\includegraphics[angle=270]{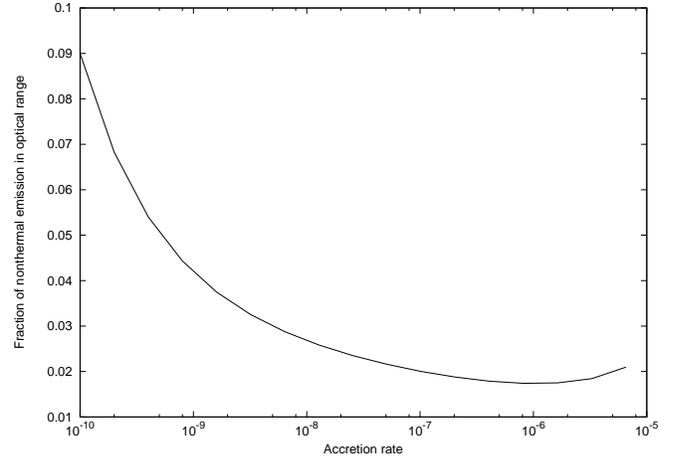}} \par}
\caption{Contribution of nonthermal component to the optical emission
$L_{\rm nt}^{\rm opt}/L^{\rm opt}$ as a function of accretion rate}
\label{fig_nonthermal_fraction_optics}
\end{figure}

A significant amount of nonthermal synchrotron emission also falls in
the optical range  (see Table \ref{table_luminosity} and
Fig.\ref{fig_nonthermal_fraction_optics}). This means that short flares
like that shown in Fig.\ref{fig_flare_lightcurve} may be detected in
observations with a 1 $\mu$s time resolution at the 6 meter telescope in
the framework of the MANIA experiment (Beskin et al. \cite{beskin_1997}).

\section{Discussion}


The analysis of existing data on possible black hole masses and
velocities is performed in comparison with the interstellar medium
structure. It is shown that in the majority of cases in the Galaxy
($>90\%$), the accretion rate $\dot m = \dot M c^2/L_{edd}$ cannot exceed
$10^{-6}-10^{-7}$ (see Fig.\ref{fig_accretion_rate}). The basis of our
analysis is the assumption of energy equipartition in the accretion
flow of Shvartsman (\cite{shvartsman_1971}). Currently, there have been no
successful attempts to build a magnetized plasma accretion theory
without this assumption -- the work of Kowalenko \& Melia
(\cite{kowalenko}) led to unphysical results, while Scharlemann
(\cite{scharlemann}) has some mathematical errors in
computing the back-reaction of the magnetic field on the matter.

The accreting plasma is initially collisionless, and it remains so until
the event horizon. The electron-electron and electron-ion free
path $\lambda\sim2.4\cdot10^3T^2n^{-1}$ even at the capture radius is
as high as $\sim10^{12}$ cm. Only the magnetic fields trapped in
plasma (the proton Larmor radius at $r_g$ is 10 cm) make it possible to
consider the problem as a quasi-hydrodynamical one; it is only due to
the magnetic field that the particle's momentum is not conserved, allowing
particles to fall towards the black hole. In addition, the
magnetic field effectively "traps" particles in a "box" of variable
size, which allows us to consider its adiabatic heating during the
fall; a correct treatment of such a process shows that for magnetized
plasma such heating is 25\% more effective than for ideal gas (see
Sec.\ref{sec_adiabatic}). Therefore, the plasma temperature in the
accretion flow grows much faster and electrons become relativistic earlier
--  $R_{rel}\approx6000$ in contrast to $R_{rel}\approx1300$ in 
Bisnovatyi-Kogan \& Ruzmaikin (\cite{bisnovatyi_1974}) and
$R_{rel}\approx200$ in Ipser \& Price (\cite{ipser_1982}). The
accretion flow is much hotter, and our estimation of "thermal"
luminosity 
\begin{equation}
L = 9.6\cdot10^{33} M_{10}^3 n_1^2 (V^2+c_s^2)_{16}^{-3}  \mbox{ erg/s}
\end{equation}
is significantly higher than those of Ipser \& Price (\cite{ipser_1982})
\begin{equation}
L_{\rm IP} = 1.6\cdot10^{32} M_{10}^3 n_1^2 (V^2+c_s^2)_{16}^{-3} \mbox{ erg/s}
\end{equation}
and Bisnovatyi-Kogan \& Ruzmaikin (\cite{bisnovatyi_1974})
\begin{equation}
L_{\rm BKR} = 2\cdot10^{33} M_{10}^3 n_1^2 (V^2+c_s^2)_{16}^{-3} \mbox{ erg/s, }
\end{equation}
while the optical spectral shape is nearly the same. 

We considered dissipation of the magnetic energy in the turbulent
current sheets (Pustilnik \cite{pustilnik_1997}) as a mechanism
supporting equipartition. In this process, the electrons  are ejected
with plasma from the current sheet and are accelerated. These electrons
have a power-law energy distribution and its emission spectrum is flat up
to the gamma band (Figs.\ref{fig_spectrum_decomposed} and
\ref{fig_spectrum_multi}). An important property of the nonthermal
emission is its flaring nature -- the electron ejection process is
discrete; typical light curves of single beams are shown in Fig.
\ref{fig_flare_structure} and \ref{fig_flare_lightcurve}.

Table \ref{table_luminosity} summarizes the observable parameters of accreting
black holes.

The black holes have significant luminosity only when they are located
in the warm hydrogen regions that occupy about 50$\%$ of the Galaxy
volume (McKee \& Ostriker \cite{mckee}). In such cases the
dimensionless accretion rate lies in the 10$^{-8}$ -- 10$^{-6}$ range,
and the bolometric luminosity of the accretion flow is
$3\cdot10^{28}$-10$^{33}$ erg/s, depending on the BH velocity and the
fraction of magnetic energy transformed into the motion of electrons
(see Table \ref{table_luminosity}). As a result, the black hole at a 100
pc distance (a sphere with this radius must contain several tens of such objects,
see Beskin et al (\cite{beskin_2000}) and Agol \& Kamionkowski
(\cite{agol_2002b})) looks like a 15-25$^{\rm m}$ optical object (due
to the "thermal" spectral component) with a strongly  variable companion
in high-energy spectral bands ("nonthermal" component). The hard
emission consists of flares, the majority of which
are generated inside a 5$r_g$ distance from the BH (see  Figs.
\ref{fig_flare_structure},\ref{fig_flare_lightcurve},\ref{fig_flare_amplitude}).
These events have the durations $\sim r_g/c$ ($\sim$ 10$^{-4}$ s),
a rate of 10$^{3}$-10$^{4}$ flares per second, and an amplitude of 2\%-6\%.
As it is seen from Table \ref{table_luminosity},
the BH variable X-ray emission may be detected by modern space-borne
telescopes.

Optical emission
consists of both a quasistationary "thermal" part and a low-frequency tail
of nonthermal flaring emission. The rate and duration of optical
flares are the same as X-ray ones, while their amplitudes are
significantly smaller. Indeed, the contribution of nonthermal component
to the optical emission (see Fig.\ref{fig_nonthermal_fraction_optics}) is
approximately $2\cdot10^{-2}$ for $\dot m = 10^{-8} - 10^{-6}$, so the mean
amplitudes of optical flares are 0.04\%-0.12\%, while the peak ones may be 1.5-2
times higher and reach 0.2\%.  Certainly, it is nearly impossible to
detect such single flares, but their collective power reaches 18-24$^{\rm
m}$ (see Table \ref{table_luminosity}) and thus may be detected in 
observations with high time resolution
($<$ 10$^{-4}$ s) by the large optical telescopes.

Our separation of the accretion flow
emission into stationary "thermal" and flaring "nonthermal" components
is a rough approximation made for the estimation of the
qualitative picture of BH observational appearances. The real behaviour
of the accreting plasma may be much more complicated.

The accretion flow is a complex dynamical system with
nonlinear feedback. This is ensured by the plasma oscillations
generated in each reconnection event, with beams of accelerated
electrons and clouds of magnetized plasma ejected from current
sheets. All these agents may act as triggers for already "prepared"
inhomogeneities which turn on magnetic energy dissipation processes.
This situation seems to be similar to the Solar one which determines its
flaring activity, and also to the case of UV Cet stars and maybe
accretion disks of X-ray binaries and active galactic nuclei. All these
non-stationary processes are characterized by power-law scalings of
flare energies with similar slopes of 1.5-2 at a very wide range of
energies -- from $10^{23}$ erg/s for the Sun to $10^{45}$ erg/s for
quasars. The universality of these processes may be interpreted in the
framework of a fractal approach as done by Bak, Tang \&
Weisenfeld (\cite{bak}), Lu \& Hamilton (\cite{lu_1991}), Lu et al.
(\cite{lu_1993}), Anastasiadis, Vlahos \& Georgoulis
(\cite{anastasiadis}), Kawaguchi \& Mineshige (\cite{kawaguchi}),
Pustilnik (\cite{pustilnik_1997}). This means the realization (at least
in active phases) of some collective state, sometimes called
"self-organized criticality" (Bak, Tang \& Weisenfeld \cite{bak})),
which is characterized by the same behaviour of the parameters on all
scales. These are percolation processes. There is evidence that
accretion flow is in this state, and so its observational appearance
(at least that related to non-stationary processes) may be predicted
and interpreted in the framework of this approach. Initial steps in
this direction have been made by Beskin \& Karpov
(\cite{beskin_2002a}), but need to be refined.

\begin{figure}
{\centering \resizebox*{1\columnwidth}{!}{\includegraphics[angle=270]{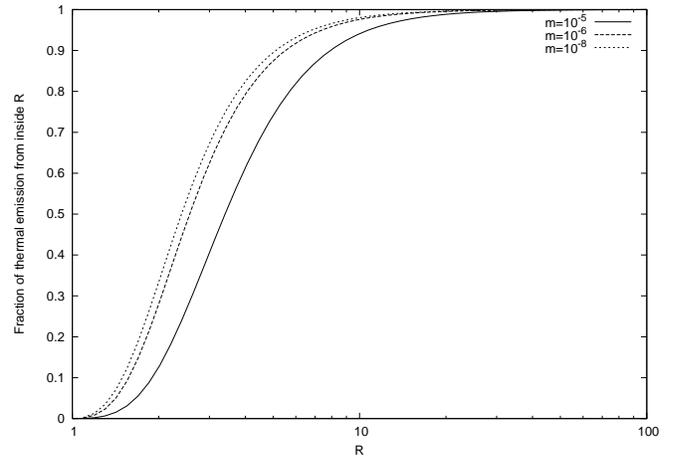}} \par}
\caption{Fraction of thermal synchrotron emission that comes from inside a given radius $R$}
\label{fig_luminosity_radial}
\end{figure}

\section{Conclusions}

During recent years the number of works dealing with single
stellar-mass black holes has significantly increased. Some are
purely theoretical (Punsly
\cite{punsly_1998a,punsly_1998b}; Gruzinov \& Quataert \cite{gruzinov};
Abramowicz et al. \cite{abramowicz}) and other provide discussions of
their observational detection (Heckler \& Colb \cite{heckler}; Fujita
et al. \cite{fujita}; Beskin et al. \cite{beskin_2000}; Agol \&
Kamionkowski \cite{agol_2002b}; Chisholm et al. \cite{chisholm}). The
importance of experiments in strong gravitational fields has been
recently noted by Damour (\cite{damour}). Kramer et al.
(\cite{kramer}) discussed the new possibilities of black holes
metric study by the investigation of radio pulsar -- BH binary
systems with new generation of radio telescopes.

In this work we tried to concretize physical properties
of plasma accreted onto the black hole within the classical
paradigm of equipartition of Shvartsman
(\cite{shvartsman_1971}). Assuming the discrete nature of the
magnetic energy dissipation processes in current sheets allows us to
clarify the shape of the synchrotron spectrum of the accretion flow. A hard
highly non-stationary nonthermal spectral component appears as an
emission of accelerated particles.  The beams accelerated in the
current sheets can generate very short flares, providing information
about the neighborhood of the event horizon (Fig.\ref{fig_flare_lightcurve}). On
the other hand it is clear from Fig.\ref{fig_luminosity_radial} that at
low accretion rates a significant amount of thermal synchrotron
radiation is generated inside $3r_g$ -- this means that the behaviour of this
component will reflect the properties of space-time in strong gravitational
fields too.

It is clear that the search for a black hole strategy may be modified in
accordance with such results. Optical high time resolution studies of
X-ray sources may be very important. Single black holes may be
contained inside the known stationary gamma sources (Gehrels et al.
\cite{gehrels}) as well as objects causing long microlensing events
(Paczynski \cite{paczynski}). Thus it is very important to  look for
X-ray emission as well as for fast optical variability of these
objects. Sample observations of the longest microlensing event
MACHO 1999-BLG-22 (Bennett et al. \cite{bennett_2001}), a stellar-mass black hole candidate,
have been performed
at the Special Astrophysical Observatory of RAS in the framework of the
MANIA experiment in 2003-2004  (Beskin et al. \cite{beskin_2005}).

The best evidence will be provided by the synchronous high time
resolution observations in optical and X-ray ranges.

Detection of the event horizon
signatures cannot result from statistical studies. A detailed
study of each object is needed to detect its specific appearance.

\begin{acknowledgements}
This work has been supported by the Russian Foundation for Basic Research
(grants No. 01-02-17857 and 04-02-17555) and by the grant in the framework of
the CNR (Italy) -- RAS (Russia) agreement on scientific collaboration. 
We also thank T.Tupolova for help in manuscript preparation.
S.K. thanks the Russian Science Support Foundation for support.
G.B. thanks the Cariplo Foundation for the scholarship and Merate Observatory for
hospitality. The Authors thank Dr. L.Pustilnik for exciting discussions.
The authors thank the anonymous referee for the valuable comments.

\end{acknowledgements}

\appendix
\section{Moments of single accelerated beam energy distribution}
\label{appendix_beam_moments}

For the case of the electron beam ejected from the current sheet at some
radius $R_0$ and free-falling to $R$ not interacting with the background
particles and beams ejected below the moments of the energy distribution may
be written using (\ref{eqn_gamma_evolution_solution}) as
\begin{equation}
\overline \gamma_b(R)= \left(\int\limits_1^{\infty}f(R_0,\gamma_0)d
\gamma_0\right)^{-1}\int\limits_1^{\infty}\frac{f(R_0,\gamma_0)\gamma_0d
\gamma_0}{C_1(R,R_0)\gamma_0+C_2(R,R_0)}
\end{equation}
\begin{equation}
\overline \gamma^2_b(R)= \left(\int\limits_1^{\infty}f(R_0,\gamma_0)d
\gamma_0\right)^{-1}\int\limits_1^{\infty}\frac{f(R_0,\gamma_0)\gamma_0^2d
\gamma_0}{C_1(R,R_0)\gamma_0+C_2(R,R_0)}
\end{equation}

For the initial distribution of the form (\ref{eqn_distribution_ejected})
this may be evaluated as
\begin{equation}
\overline \gamma_b(R) = \Gamma(R_0)\Fcurve_1\left(\sqrt{\frac{R}{R_0}},
\frac{A\Gamma(R_0)}{R}\left(1-\left[\frac{R}{R_0}\right]^{3/2}\right)\right)
\label{eqn_mean_gamma}
\end{equation}
\begin{equation}
\overline \gamma^2_b(R) = \Gamma^2(R_0)\Fcurve_2\left(\sqrt{\frac{R}{R_0}},
\frac{A\Gamma(R_0)}{R}\\
\left(1-\left[\frac{R}{R_0}\right]^{3/2}\right)\right)\\
\label{eqn_mean_gamma2}
\end{equation}
where new functions
\begin{equation}
\begin{array}{lll}
\Fcurve_1(A,B)&=&\int\limits_0^{\infty}\frac{x e^{-x}dx}{Ax+B}\\
&=&\frac1{A^2}
\left(A-B \cdot\exp{\left(\frac{B}{A}\right)}\mbox{Ei}_{1}\left(\frac{B}{A}
\right) \right)\\
\end{array}
\label{eqn_fcurve}
\end{equation}
\begin{equation}
\begin{array}{lll}
\Fcurve_2(A,B)&=&\int\limits_0^{\infty}\frac{x e^{-x}dx}{(Ax+B)^2}\\
	      &=&\frac1{A^3}
\left (-A+(1+B)
\cdot\exp{\left (\frac{B}{A}\right)}\mbox{Ei}_{1}\left (\frac{B}
{A}\right ) \right ) \\
\label{eqn_fcurve2}
\end{array}
\end{equation}
are introduced and an expression for the integral exponent
\begin{equation}
\mbox{Ei}_n(x)=\int\limits_1^{\infty}\frac{e^{-\xi x}d\xi}{\xi^n} .
\end{equation}
is used.

\end{document}